\documentclass[%
 reprint,
 amsmath,amssymb,
 aps,
]{revtex4-2}

\usepackage{hyperref}
\usepackage{graphicx}
\usepackage{dcolumn}
\usepackage{bm}
\usepackage[utf8]{inputenc}
\usepackage{float}
\usepackage{amssymb}
\usepackage[T1]{fontenc}
\usepackage{color,soul}
\usepackage{mathtools}
\usepackage{epsfig}
\usepackage{color}
\usepackage{url}
\usepackage{siunitx}
\usepackage{enumitem}
\usepackage{caption}
\usepackage{amsmath}
\usepackage{subfig}
\usepackage{bm}
\usepackage{mathptmx}

\begin{document}

\preprint{PRB}

\title[Numerical Model Of Harmonic Hall Voltage Detection]{Numerical Model Of Harmonic Hall Voltage Detection \\ For Spintronic Devices}

\author{Sławomir Ziętek$^{ a}$}
    \email{zietek@agh.edu.pl}

\author{Jakub Mojsiejuk$^{ a}$}%
    \email{mojsieju@agh.edu.pl}
\author{Krzysztof Grochot$^{ a,b}$}

\author{Stanisław Łazarski$^{ a}$}

\author{Witold Skowroński$^{ a}$}

\author{Tomasz Stobiecki$^{ a,b}$}

 \affiliation{a) AGH University of Science and Technology, Institute of Electronics, Al. Mickiewicza 30, 30-059 \\
 b) AGH University of Science and Technology, Faculty of Physics and Applied Computer Science, Al. Mickiewicza 30, 30-059 Krak\'{o}w, Poland}

\date{\today}

\begin{abstract}
We present a numerical macrospin model for harmonic voltage detection in multilayer spintronic devices. The core of the computational backend is based on the Landau-Lifshitz-Gilbert-Slonczewski equation, which combines high performance with satisfactory, for large-scale applications, agreement with the experimental results.
We compare the simulations with the experimental findings in Ta/CoFeB bilayer system for angular- and magnetic field-dependent resistance measurements, electrically detected magnetisation dynamics, and harmonic Hall voltage detection. Using simulated scans of the selected system parameters such as the polar angle $\theta$, magnetisation saturation ($\mu_\textrm{0}M_\textrm{s}$) or uniaxial magnetic anisotropy ($K_\textrm{u}$) we show the resultant changes in the harmonic Hall voltage, demonstrating the dominating influence of the $\mu_\textrm{0}M_\textrm{s}$ on the first and second harmonics. In the spin-diode ferromagnetic resonance (SD-FMR) technique resonance method the ($\mu_\textrm{0}M_\textrm{s}$, $K_\textrm{u}$) parameter space may be optimised numerically to obtain a set of viable curves that fit the experimental data.

\end{abstract} 

\keywords{Landau-Lifshitz-Gilbert-Slonczewski equation, macrospin model, numerical methods, spintronic devices}
\maketitle

\section{Introduction}
Development of novel electronic devices utilising electron spin for its operation has become an increasingly important branch of science and engineering in the past decade\cite{Dieny2020, bhatti2017spintronics, ikegawa2020magnetoresistive, hirohata2020review}. Specifically, taking advantage of both the electron spin and charge creates an opportunity for further miniaturisation and increase in the energy efficiency \cite{Manipatruni2019} of the electronic devices. However, experimental investigations typically require expensive and time-consuming fabrication processes as well as a unique measurement methodology. Computer-aided optimisation of spintronics devices, coupled with the prediction of their electric and magnetic properties, vastly reduces the number of experimental iterations and allows a faster and more efficient prototype device development. In addition, modelling of multilayer devices enables the extraction of parameters that are typically hard to obtain from experiments.

After the experimental discovery of the so-called spin-orbit torque (SOT) \cite{ralph_spin_2008, brataas2012current, manchon2019current, zhang2021field, chen2020manipulation, liu2012spin, song2021spin} there have been numerous studies on spin current generation in nonmagnetic materials with high spin-orbit coupling additionally improved by interfacial effects \cite{skowronski2019determination, ogrodnik2021study}. Utilising SOT may lead to fast magnetisation switching\cite{grimaldi2020single} and more durable magnetic memory design \cite{zhou2020design}. To quantify the efficiency of the effect, typically called the spin-Hall angle, one computes the ratio of the spin current to the charge current. 
In recent years, there has been a sprout in the development of spin-Hall angle measurement techniques such as spin-torque ferromagnetic resonance (ST-FMR)\cite{liu2011spin}, magnetisation switching induced by current \cite{hao2015giant} and harmonic Hall voltage detection\cite{hayashi_quantitative_2014, kim2013layer}.
The latter method does not require a sophisticated fabrication protocol, nor the determination of additional thermal or high-frequency effects, and allows for the extraction of damping-like and field-like effective fields, from which the spin-Hall efficiencies may be calculated. 

In this work, we employ the SOT effect as a basis for the electrical model that simulates the harmonic Hall voltage technique using field-like (FL) and damping-like (DL) SOT torques. 
Furthermore, we present a highly efficient macrospin modelling software for electrical detection of static and dynamic magnetic properties in multilayer spintronic devices. Our model demonstrates good agreement with the experimental data and provides additional insights into different aspects of magnetisation dynamics and harmonic Hall measurement. Moreover, we show the dependence of magnetisation saturation, and anisotropy on the harmonic Hall voltage detection, with the former having a much stronger impact on the final result. 
The model package, called CMTJ (C++ Magnetic Tunnel Junctions), is provided in both C++ and Python interfaces, along with the hereby described postprocessing steps. As a demonstration of the ease of use and the speed of the model, all simulations conducted in this article may be reproduced within half-an-hour on a modern laptop. The simulation scripts along with the simulation package itself are open sourced under this address: 
\url{https://github.com/LemurPwned/cmtj}.

\section{Numerical model of electrical detection}
\subsection{Theoretical background}
\label{sec:eff-torq}
First, we present a theoretical model of the magnetisation dynamics together with the electrical detection methodology. We adapt the standard Landau-Lifshitz-Gilbert-Slonczewski (LLGS) equation, like presented in, e.g., Nguyen et. al\cite{nguyen_spinorbit_2021}, into LL-form\cite{ralph_spin_2008}, such that it may be implemented in the numerical engine. The LLGS equation itself is given in the form
\begin{multline}
	  \frac{\textrm{d}\textbf{m}}{\textrm{dt}} = -\gamma_0 \textbf{m} \times \textbf{H}_{\mathrm{eff}} + \alpha_\textrm{G} \textbf{m}\times \frac{\textrm{d}\textbf{m}}{\textrm{dt}} \\
	 -\gamma_0|H_\textrm{FL}|(\textbf{m} \times  \textbf{p}) -\gamma_0|H_\textrm{DL}|(\textbf{m}\times\textbf{m}\times \textbf{p})
\label{eq:llg-sot}
\end{multline}    
where \(\textbf{m} = \frac{\textbf{M}}{\mu_\textrm{0}M_\textrm{s}}\) is the normalised magnetisation vector with $\mu_\textrm{0}M_\textrm{s}$ as the magnetisation saturation, $\alpha_\mathrm{G}$ is the dimensionless Gilbert damping parameter, $\textbf{H}_\mathrm{eff}$ is the effective field vector, $\textbf{p}$ is the polarisation vector, and \(\gamma_{0}\) is the gyromagnetic factor. The terms representing the magnitude of the $H_\mathrm{FL}$ (field-like) and $H_\mathrm{DL}$ (damping-like) torque fields have a concrete connection with the spin-Hall angle and correspond to the DL and FL SOT. See the Appendix for more details on the transition from the LLGS form to the numerically viable LL form. 
The effective field vector $\textbf{H}_\mathrm{eff}$ is usually composed of various field contributions, which, depending on the context of the simulation, may be added or disabled.
The simulation package in question already provides a range of such contributions, including the interlayer exchange coupling (IEC), dipole and demagnetisation interactions, magnetic anisotropy, and external magnetic field contribution. In the experimental data, presented in the following sections, we investigate a bilayer (heavy metal/ferromagnet) structure, thus we may omit IEC and dipole interactions in our simulations.
For numeric integration, we employ the higher order Runge-Kutta method. After each integration step, the $\textbf{m}$ vectors are normalised to avoid cumulative numerical error. In deciding on the value of integration step we usually compromise the computation time and the stability of the numerical solution. For majority of simulations this parameter is in range of a femto-second or lower. As a good rule of thumb, it is better to start at a lower value (around $10^{-13}\si{s}$) and steadily increase it, verifying that the obtained results are still consistent.

\subsection{Modelling of magnetoresistance effects}

\begin{figure}
    \centering
    \includegraphics[scale=0.25]{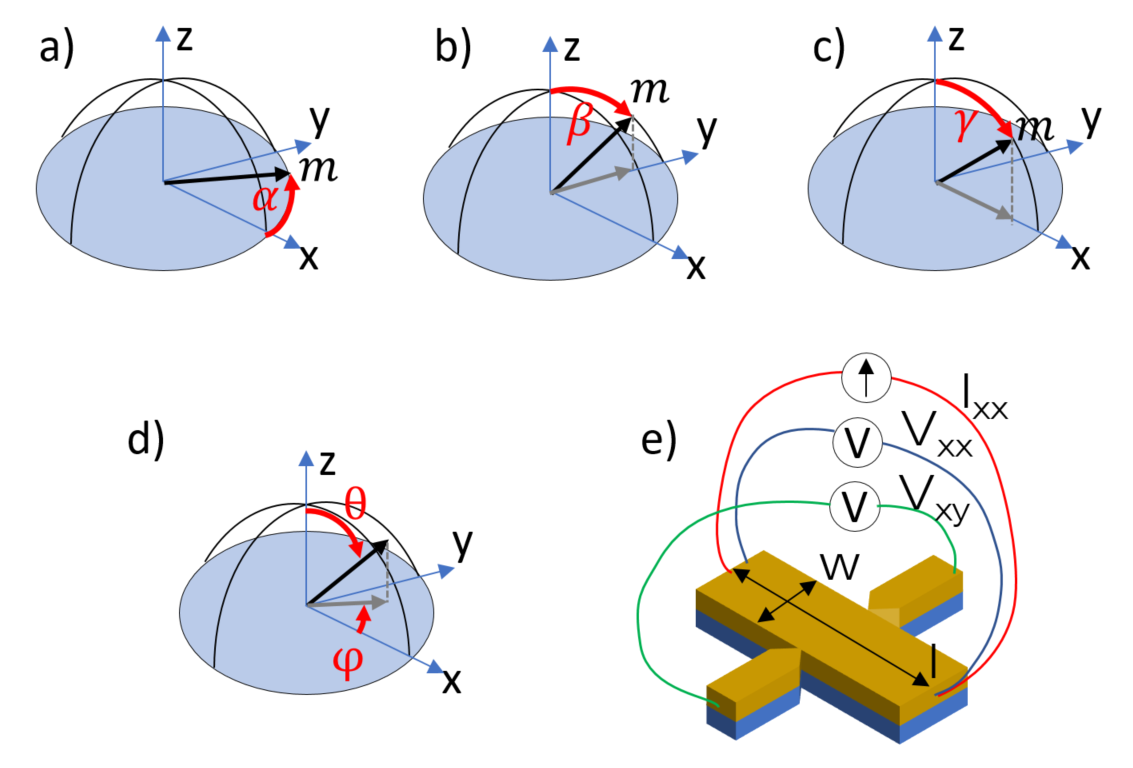}
    \caption{The reference diagram for measurement arrangements used in this work. Fig.(a-c) display the angles $\alpha, \beta, \gamma$ used in simulations and experiment in the saturating field. The angle $\theta$ is taken to be the polar angle and $\phi$ is the azimuthal angle (d). Fig.(e) depicts the application of the voltmeter in the measurement.}
    \label{fig:coordinates}
\end{figure}
In our simulations, we can easily compute longitudinal ($R_\mathrm{xx}$) and transverse ($R_\mathrm{xy}$) magnetoresistance loops for current in-plane (CIP) and current perpendicular to the plane (CPP) configurations (see Fig.\ref{fig:coordinates} d-e), as a function of the magnetic field magnitude or angle. To obtain $R_\mathrm{xx}$ and $R_\mathrm{xy}$ loops we adopt a model from \cite{kim2016}, given by Eq.\ref{eq:Rxx} - \ref{eq:Rxy}:
\begin{equation}
    \label{eq:Rxx}
    R_\mathrm{xx} = R_\mathrm{xx0} + (\Delta R_\mathrm{AMR} m_x^2 + \Delta R_\mathrm{SMR} m_y^2)
\end{equation}
\begin{equation}
    \label{eq:Rxy}
    R_\mathrm{xy} = R_\mathrm{xy0} + \frac{1}{2}\Delta R_\mathrm{AHE} m_z + \frac{w}{l}(\Delta R_\mathrm{SMR} + \Delta R_\mathrm{AMR})m_xm_y
\end{equation}
where $\Delta R_\mathrm{AMR}$ and $\Delta R_\mathrm{SMR}$ are the magnitudes of the anisotropic (AMR) and spin-Hall (SMR) magnetoresistances (both in $\Omega$) in $R_\mathrm{xx}$ configuration. $\Delta R_\mathrm{AHE}$ is the magnitude of Anomalous Hall Effect (also in $\Omega$), $l$ is the length and $w$ is the width of the sample. In the dynamic state, the resistance is calculated as a function of time and then used for calculation ST-FMR and harmonic Hall voltages.

\subsection{Spin Diode Ferromagnetic Resonance (SD-FMR)}
\label{sec:sd}
In addition to the inductive and optical magnetometry methods, SD-FMR has proven to be a powerful experimental tool in the study of magnetisation dynamics in microwave spintronic devices such as oscillators or detectors \cite{Locatelli2014}. When the alternating current (AC) is passed through a magnetoresistive element, it generates small changes of magnetisation driven by secondary effects of Oersted field, spin transfer torque (STT) or SOT, which finally lead to the oscillations of resistance \cite{Tulapurkar2005}. Mixing of the AC and oscillating resistance gives raise to the mixing voltage which has both direct current (DC) and AC components at first ($f'=f$) and second ($f'=2f$) harmonic frequency. Those components may be extracted with different filters in postprocessing step. For SD-FMR, this DC component is called $V_\mathrm{DC}$ voltage and in the experiment it is separated from AC components using a bias-T filter \cite{sankey2008measurement, ziketek2015rectification}.

In the simulation setup, we compute $V_\mathrm{DC}$ analogously to the experiment: a sinusoidal current $I_\mathrm{RF}$ with microwave frequency $f$ is applied along $x$ axis, which in turn generates a tangent Oersted field $H_\mathrm{Oe}$. This guides oscillations of the magnetisation and  time-varying resistance, $R_\mathrm{xx}(t)$ and $R_\mathrm{xy}(t)$. Multiplying the sinusoidal current excitation and the resistance produces a voltage $V(t)$, that akin to experiment, has the DC, AC components. To separate the DC component we filter the voltage signal with a digital low pass filter (LPF). Taking a mean of the LPF-filtered voltage signal yields one $V_\mathrm{DC}$ value for each magnetic field and frequency. 

\subsection{Harmonic Hall voltage detection}
We determine the spin torque components with harmonic Hall voltage measurements in the low frequency regime. The established methods use either magnetic field dependence \cite{hayashi_quantitative_2014} or angular dependence \cite{avci2014} to analyse the first and second harmonic Hall voltage signals under AC (below the resonance frequency) excitation. CMTJ simulates these two approaches by extending the previously described SD-FMR method with a simple computation of phase and amplitude in the first and second harmonic, for $R_\mathrm{xy}$ configuration. For low-frequency regimes, when the magnetisation vector does not undergo large-angle variations, the results happen to be sufficiently close. 

In our simulations, we follow the SOT formulation as described in Eq.\ref{eq:llg-sot}. At a low frequency, below the resonance, we perform a field scan, where at each field step the system is excited with a sinusoidal torque signal, with separate amplitudes for damping- and field-like torques. We compute the $R_\mathrm{xy}$ magnetoresistance using Eq.\ref{eq:Rxy}, and calculate the Fast Fourier transform (FFT) of the mixing voltage signal to obtain the amplitude and phase. We remove the offset from the experiment and simulation data for the first harmonic, and we convert the second harmonic phase from radian to voltage. Then, following the lock-in\cite{lockin} operation, we first take the cosine of the simulated phase and then multiply by the amplitude of the signal at that second harmonic. In such a way, we obtain the amplitudes consistent with what we get from the experiment, while keeping all simulation parameters realistic (layer parameters, applied current density, and torques come well within the range observed during the experiment measurement).

\section{Results}
We now turn to a comparison of numerical simulations conducted using CMTJ with the experimental results obtained on the Ta(5)/CoFeB(1.45)/MgO(2)/Ta(1) structure (thickness in nanometers). The system has been patterned into Hall-bars enabling both static and low-frequency longitudinal and transverse resistance measurements as well as magnetisation dynamics characterisation using SD-FMR technique. Ta underlayer was chosen such that it generates significant SOT and SMR\cite{cecot2017influence}. The selected CoFeB thickness results in net perpendicular anisotropy induced by the dominating interfacial anisotropy component. For low-frequency $R_\textrm{xx}$ and $R_\textrm{xy}$ measurements, the excitation voltage was fixed to 1V. High frequency measurements were performed with the radio frequency (RF) signal of power P = 16 dBm. The details of the sample fabrication are presented in Ref.\cite{lazarski2021}.

The process of numerical harmonic detection is composed of several steps. Firstly, we compute the magnetoresistance parameters that will serve as a basis for our further simulations. In particular, we use our model to fit the magnetoresistance parameters using the angular dependencies of the resistance in the saturating magnetic field. This permits us to determine the resistance values: AMR, SMR, AHE. Then, we obtain the magnetisation, saturation and magnetic anisotropy from the R-H loops. In the next step, we simulate the SD-FMR maps for $R_\mathrm{xx}$ and $R_\mathrm{xy}$ configuration using previously determined parameters. We compare them with the dispersion relations from the SD-FMR measurements for those two electrical configurations. Finally, using all the parameters that were determined in the previous steps, we reproduce the first and second harmonics measurements.

\subsection{Resistance measurement}
\begin{figure}[h!]
    \centering
    \includegraphics[width=\linewidth]{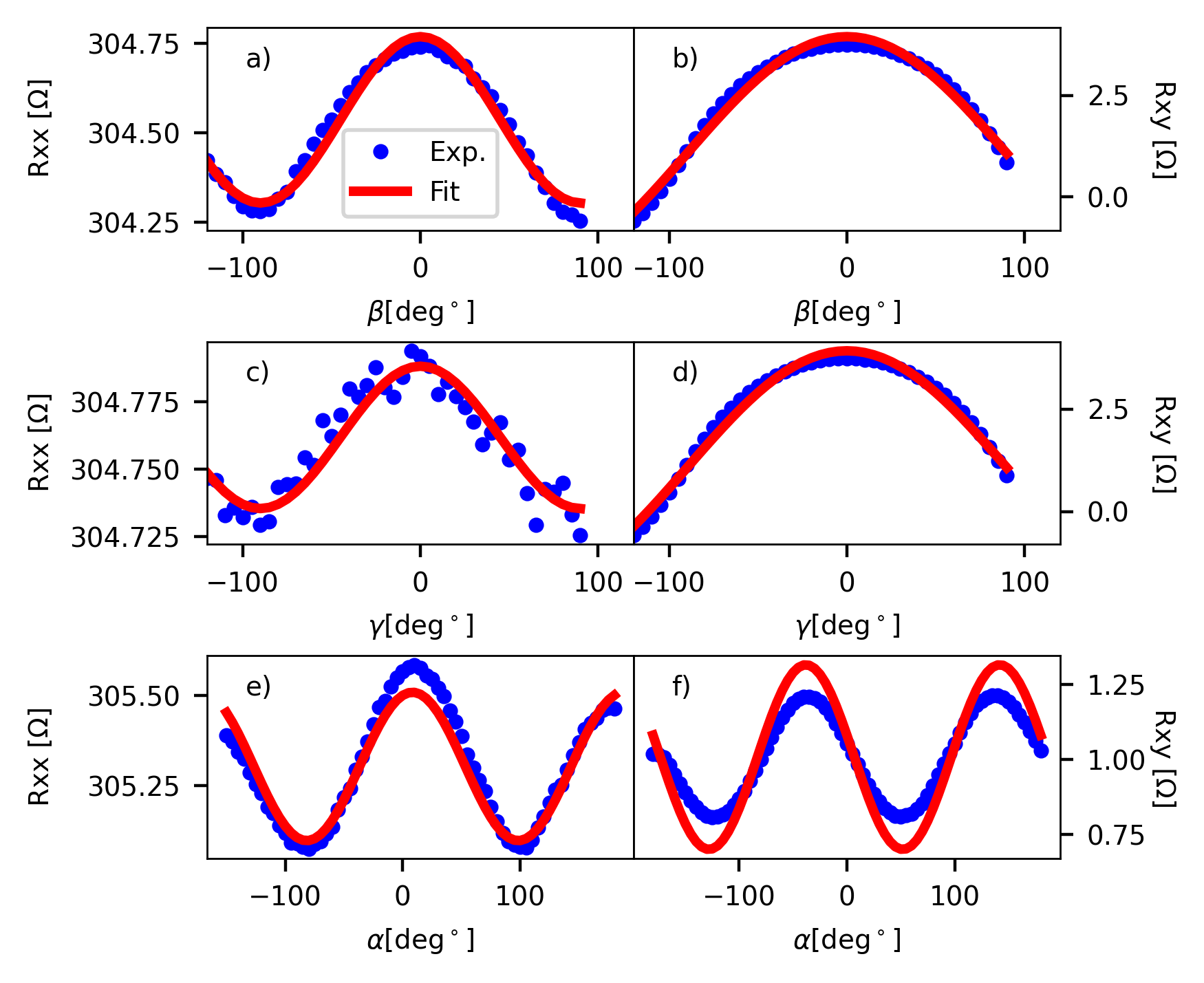}
    \caption{Comparison measurements and simulations for (a,c,e) $R_{xx}$ and (b,d,f) $R_{xy}$ at the rotations of saturating magnetic field in $\alpha$, $\beta$ and $\gamma$ angles. Experimental results are marked with blue dots, red lines represent simulation.}
    \label{fig:R-angles}
\end{figure}

To determine the magnetoresistance and magnetic parameters of the investigated sample, we perform a series of angular measurements in the saturating magnetic field and field scans at preset directions.
Angular dependencies of $R_\mathrm{xx}$ and $R_\mathrm{xy}$ measured at magnetic field of 1 T sweeping at $\alpha$, $\beta$ and $\gamma$ angles are shown in Fig.~\ref{fig:R-angles}. Red line represents the results obtained from the numerical model with electrical parameters listed in the Table \ref{tab:parameters}. A small discrepancy from a perfect sine waveform may be caused due to the FM layer being not fully saturated.
\begin{table}[h!]
\caption{Optimal parameters used in the simulations.}
\begin{ruledtabular}
\centering
\begin{tabular}{ccc}
Parameter        & Value       & Unit                   \\ \colrule
$\mu_\textrm{0}M_\mathrm{s}$            & 0.525         & T                      \\
$K_\mathrm{u}$            & 0.154      & $\si{MJ}/\si{m}^3$ \\
$\alpha_\mathrm{G}$         & 0.03        &              -          \\
$t_\mathrm{FM}$            & 1.45      & $\si{\nm}$                      \\
$\Delta R_\mathrm{SMR}$ & -0.464       & $\Omega$               \\
$\Delta R_\mathrm{AMR}$ & -0.053       & $\Omega$               \\
$\Delta R_\mathrm{AHE}$ & -5.71        & $\Omega$               \\
$w$                & 30          & $\si{\um}$                \\
$l$                & 20          & $\si{\um}$       \\
$|H_\textrm{DL}|$                & 420          & $\si{A/m}$       \\
$|H_\textrm{FL}|$                & 574          & $\si{A/m}$       
\end{tabular}
\label{tab:parameters}
\end{ruledtabular}
\end{table}

\begin{figure}[h!]
    \centering
    \includegraphics[width=\linewidth]{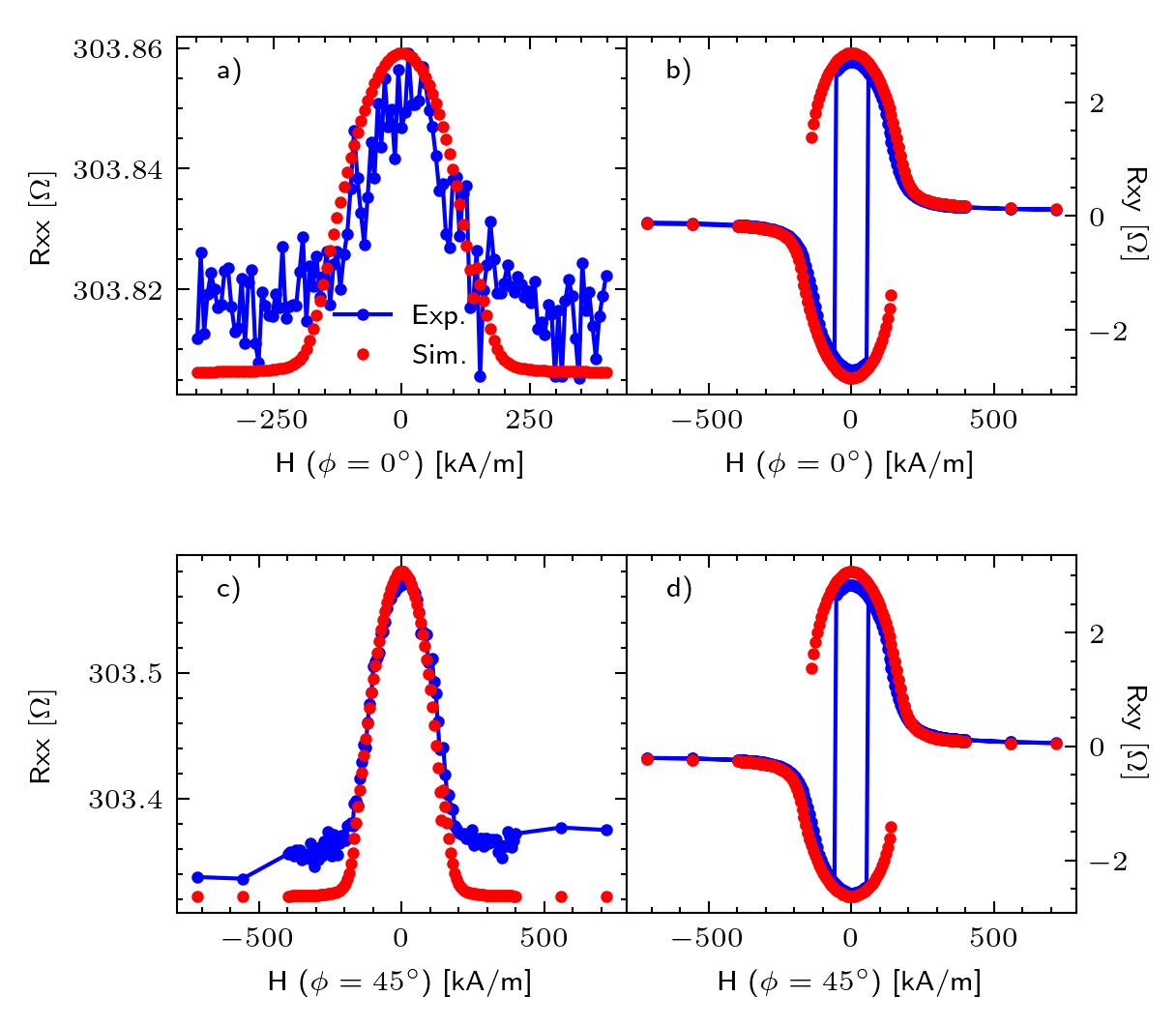}
    \caption{Comparison measurements data points and simulations for $R_\mathrm{xx}$ (a,c) and  $R_\mathrm{xy}$ (b, d) at sweeps of in plane ($\theta$ = 90$^\circ$) magnetic field at $\phi$ = 0$^\circ$ and 45$^\circ$. Blue connected dots indicate the experimental results, whereas solid red points represent simulation results obtained with CMTJ.}
    \label{fig:R-H}
\end{figure}

Fig.\ref{fig:R-H} shows experimental results of $R_\mathrm{xx}$ and $R_\mathrm{xy}$ as a function of magnetic field applied along a $\textrm{XY}$ plane at $0^{\circ}$ and $45^{\circ}$. Corresponding simulations reproduced by CMTJ are depicted as red points in the same figure. Higher switching field in the simulated loops observed for $R_\mathrm{xy}(H)$ dependency may be explained by the thermally activated magnetic domain switching\cite{czapkiewicz2008thermally} that we did not take into account in this version of the model. The following parameters reproduce experimental findings to a good degree of precision: the saturation magnetisation $\mu_0 M_\mathrm{s}$ = 0.525 T, the nominal thickness of CoFeB ($t_\textrm{FM}$) of 1.45 nm, the magnetic perpendicular anisotropy $K_\mathrm{u}$ = 0.154 MJ/m$^3$ and resistance parameters of $\mathrm{\Delta R_\mathrm{AMR}}$ = -0.053 $\mathrm{\Omega}$, 
$\mathrm{\Delta R_\mathrm{SMR}}$ = -0.464 $\mathrm{\Omega}$, $\mathrm{\Delta R_\mathrm{AHE}}$ = -5.71 $\mathrm{\Omega}$, all summarised in Table \ref{tab:parameters} for convenience.

\subsection{Magnetisation dynamics}
\begin{figure}[h!]
    \centering
    \includegraphics[width=\linewidth]{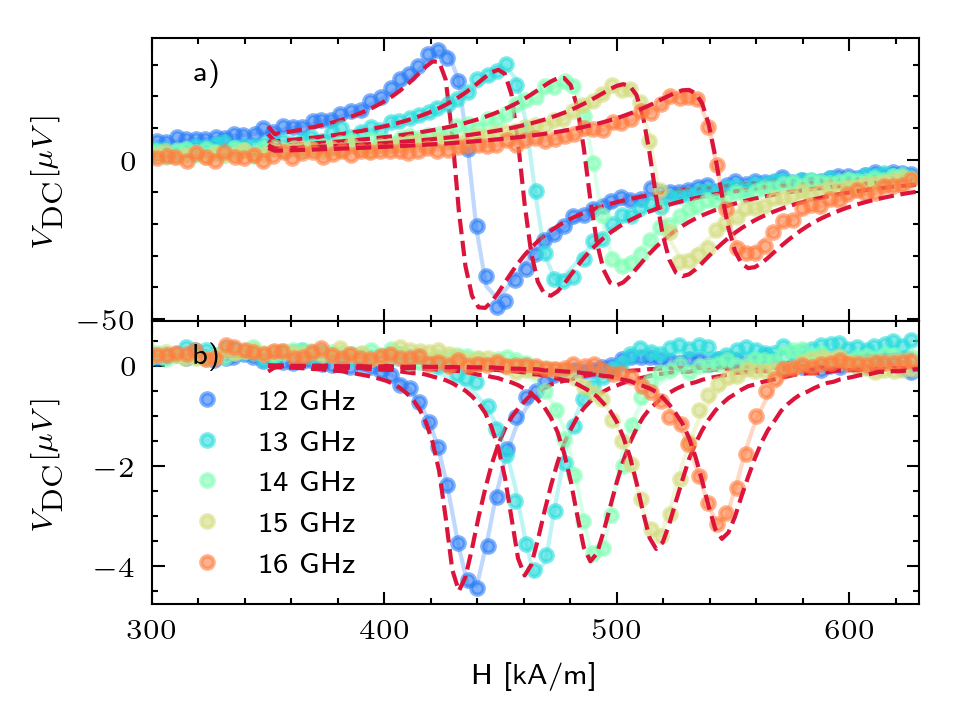}    
    \caption{Examples of $V_\textrm{DC}$ voltage measured as a function of external magnetic field applied in plane ($\theta$ = 90$^\circ$) measured at (a) $R_\textrm{xy}$ (H applied at $\phi$ = 0$^\circ$) and (b) $R_\textrm{xx}$ (H applied at $\phi$ = 45$^\circ$) electrical configuration for frequencies ranging from 12 to 16 GHz. Coloured points are experimental data, the Lorentz fits to the (\ref{eq:Harder}) are marked with solid line of the same colour. Simulations corresponding to each frequency are marked with red dashed lines, for $R_\textrm{xx}$ configuration current was $0.4\si{mA}$ while for the $R_\textrm{xy}$ one, $0.75\si{mA}$. Finally, we get the best agreement with the $\alpha_\textrm{G} = 0.03$, the other simulation parameters follow Table\ref{tab:parameters}.}
    \label{fig:vsd-lines}
\end{figure}

\begin{figure}[h!]
    \centering
    \includegraphics[width=0.9\linewidth]{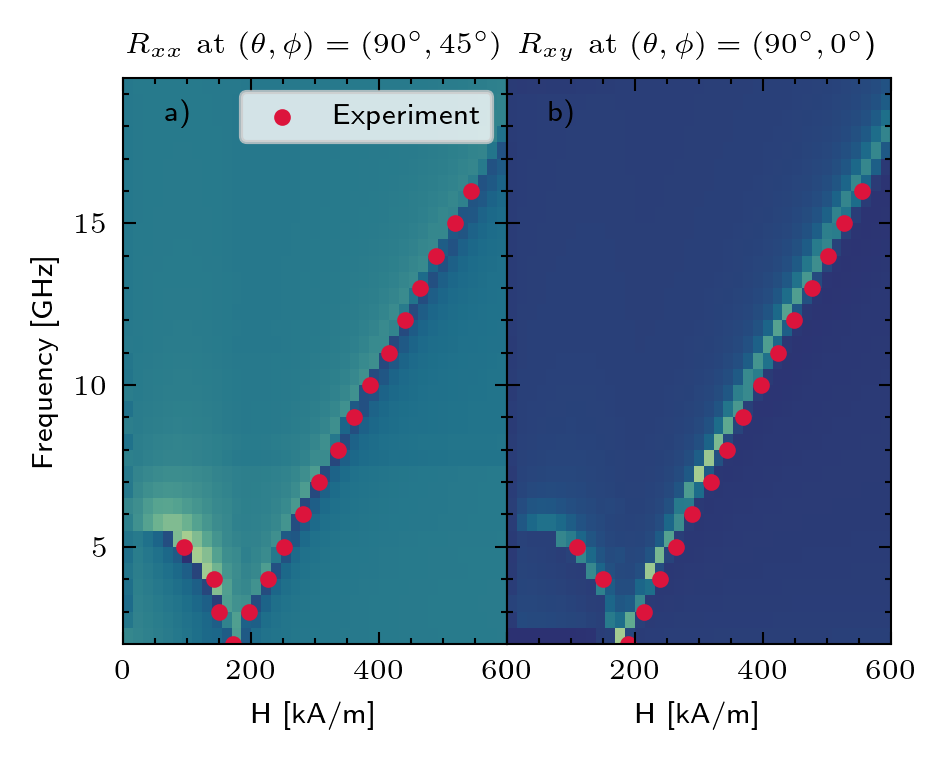}    
    \caption{Dispersion relations of SD-FMR measurements for (a) longitudinal, and (b) transverse. Red dots represent experimental data obtained for both configuration.}
    \label{fig:dynamics}
\end{figure}

Magnetisation dynamics was measured using SD-FMR technique with a fixed RF power of 16 dBm, frequency between 1 and 18 GHz, and a magnetic field swept between 0 and 600 kA/m.
An example of the measured and simulated spectra for the transverse and longitudinal magnetoresistance measurement configurations are presented in Fig.\ref{fig:dynamics}. In the Fig.\ref{fig:vsd-lines} we plot individual resonance modes of both configurations ($R_\textrm{xx}$, and $R_\textrm{xy}$) in a selected range of higher frequencies (12-16 GHz). Generally, the half-widths as well as the resonance peaks of the simulated runs (the dashed red line) remain in good agreement with the experiment marked with coloured dots. The coloured lines represent the Lorentz fit, which was computed with the following formula:
\begin{equation}
    \label{eq:Harder}
    V_{DC}(H) = A_{S}L + A_{A}D
\end{equation}
\begin{equation}
    \label{eq:LD}
    L = \frac{\Delta H^2}{(H-H_r)^2 + \Delta H^2}, \quad
    D = \frac{\Delta H(H-H_r)}{(H-H_r)^2 + \Delta H^2}
\end{equation}
where: $A_S$ and $A_A$ are amplitudes of symmetric and anti-symmetric components of the resonance line, $H_\mathrm{r}$ is the resonance field, and $\Delta H$ is the linewidth.

\subsection{Harmonic Hall detection}

\begin{figure}[h!]
    \centering
    \includegraphics[width=\linewidth]{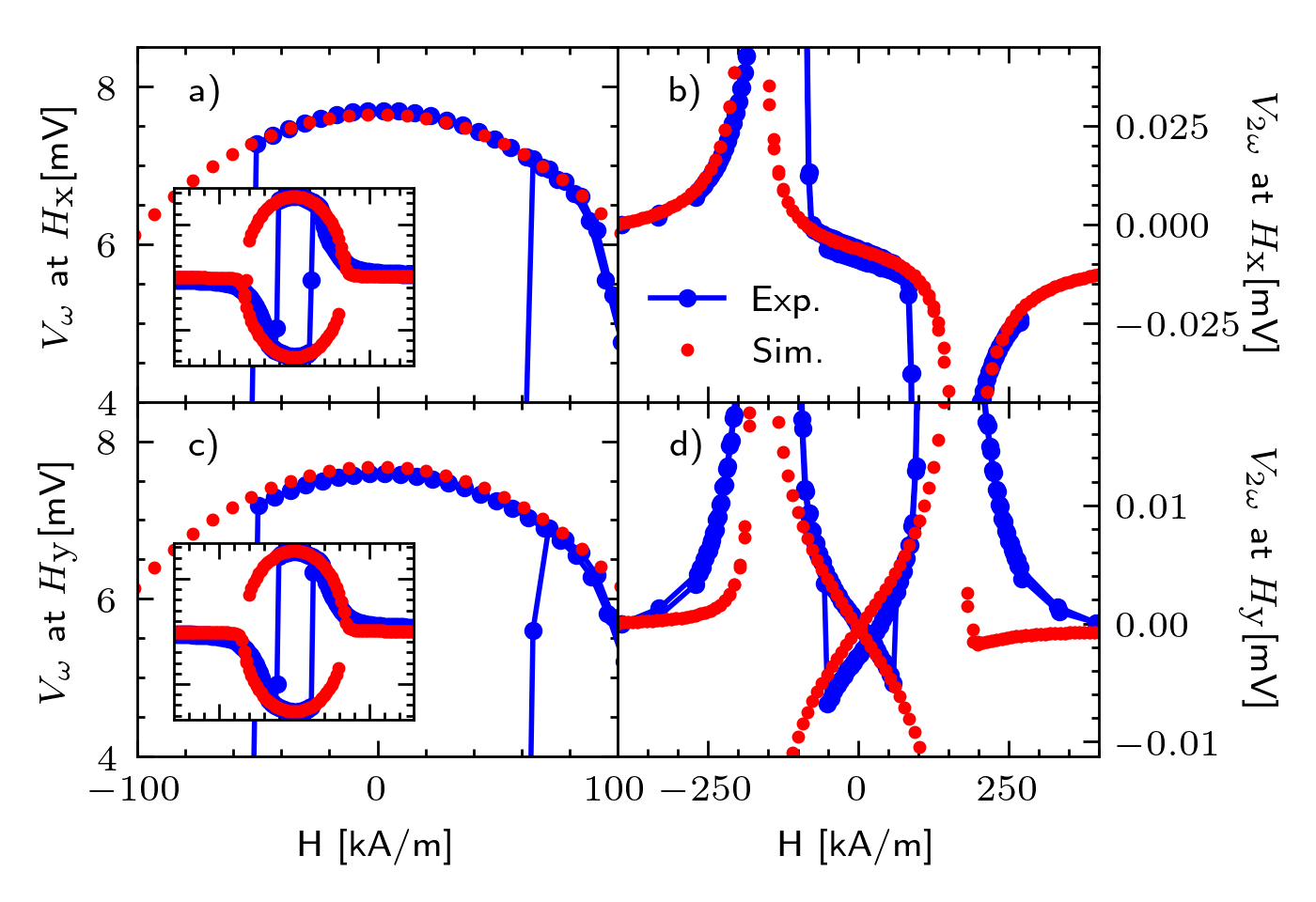}
    \caption{The best fit for first and second harmonics in the following arrangements: (a, b) $H_x$ with $\theta = 90^{\circ}, \phi = 0^{\circ}$ and (c, d) $H_y$ with $\theta = 90^{\circ}, \phi = 90^{\circ}$. Blue dots represent the experimental data, and the red dots depicts the simulation results from CMTJ. The figures (a) and (c) show the first harmonic response, while (b) and (d) demonstrate the the second harmonic. The primary fitting variables are the curvature of the quadratic region in the first harmonics and the slope in the linear section in the second harmonic curve. In the inset, the complete view of the first harmonic voltage in the field range from -400 to 400 $\si{k\frac{A}{m}}$.}
    \label{fig:harmonics-fit}
\end{figure}

\begin{figure*}[ht]
    \centering
    \includegraphics[width=0.8
    \textwidth]{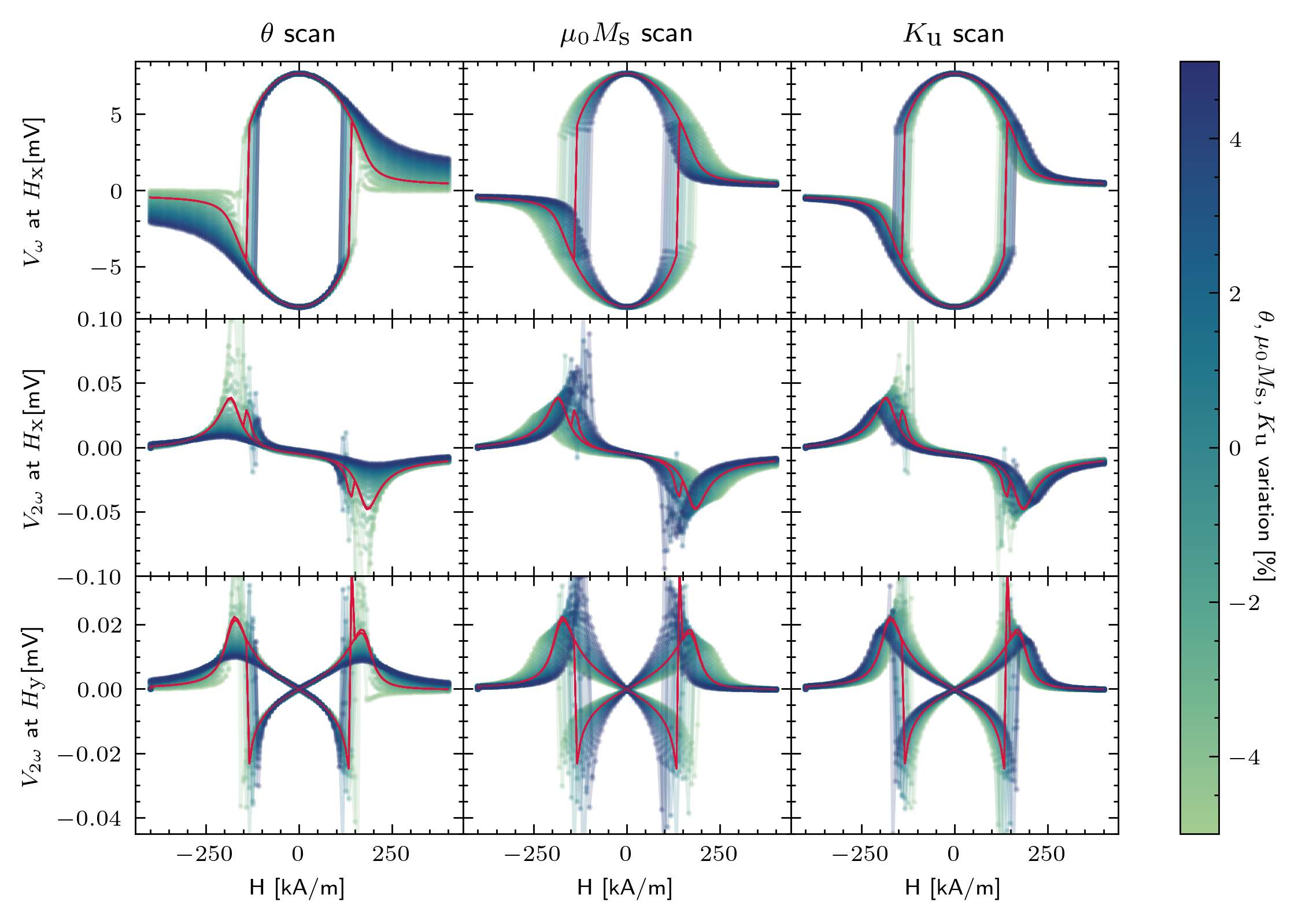}
    \caption{The figure presents influence of a parameter $\mu_\textrm{0}M_\textrm{s}$, $K_\textrm{u}$ and $\theta$ on first and second harmonic scans. Red lines represent the original curve simulated with parameters from Table\ref{tab:parameters}. We simulated the system by varying selected parameter, but keeping the rest fixed. We see that increasing $\mu_\textrm{0}M_\textrm{s}$ reveals an inverse effect to increasing $K_\textrm{u}$, but the rate of change for the latter is much smaller than the one for the former. The range of $\theta$ scan was centred around $94.8^\circ \pm 5\%$, for the $\mu_\textrm{0}M_\textrm{s}, K_\textrm{u}$ scans the default angle was $\theta=92^\circ$.}
    \label{fig:parameter-scan}
\end{figure*}

In the experimental setup, we are primarily interested in measuring the effective torque efficiencies ($\xi$), both the $\mathrm{DL}$ and $\mathrm{FL}$. Those efficiencies may be easily calculated given the values of $H_\mathrm{DL}$, and $H_\mathrm{FL}$ fields at a current density $j_e$\cite{nguyen_spinorbit_2021}:
\begin{equation}
    \xi_\mathrm{DL/FL} = \frac{2e\mu_0 \mu_\textrm{0}M_\textrm{s}}{\hbar}\frac{t_\mathrm{FM}H_\mathrm{DL/FL}}{j_e}
\label{eq:hall-angle}
\end{equation}
where $\mu_\textrm{0}M_\textrm{s}$ is the magnetisation saturation, $t_\textrm{FM}$ is the thickness of the ferromagnetic layer.
The $H_\mathrm{DL}$ and $H_\mathrm{FL}$ fields may be computed from the first $V_\omega$ and and second $V_{2\omega}$ harmonic responses in two arrangements -- longitudinal (L), later called $H_\mathrm{x}$ and transversal (T), marked $H_\mathrm{y}$. For instance, one can obtain $H_\textrm{DL}$ with the following formula:
\begin{align}
    \label{eq:torques}
    H_{\mathrm{DL}} = -\frac{2}{\zeta}\frac{\rho_L \pm 2\kappa \rho_T}{1 - 4\kappa^2}
\end{align}
where $\kappa$ is the ratio of PHE and AHE resistance and $\rho_{L/T} = \partial V_{2\omega}/\partial H_\textrm{ext}^{L/T}$ for longitudinal $L$ and transverse $T$ arrangement respectively.
In the $H_\mathrm{x}$ (longitudinal) setting, we apply the external field at $\phi=0^{\circ}$ and $\theta=90^{\circ}$, whereas for $H_\mathrm{y}$ (transverse) arrangement we have $\phi= 90^{\circ}$ and $\theta=90^{\circ}$.
The parameter $\zeta = \partial^2 V_\omega/\partial H_\textrm{ext}^2$ is obtained by fitting the low-regime region of the first harmonic, $V_\omega$ to a quadratic function. The $V_{2\omega}$ should resemble a linear function in the same low-field regime as in the case of the first harmonic. We fit that region to obtain the function slope $\rho$ for both $H_\mathrm{x}$ and $H_\mathrm{y}$ arrangements. Interchanging the subscripts $L$ and $T$ in the Eq.\ref{eq:torques} yields the value of $H_\textrm{FL}$.

The parameters $\mu_\textrm{0}M_\textrm{s}$ and $K_\textrm{u}$ were not adjusted in this step and the values of the torques were determined from the experimental data using Eq.\ref{eq:torques}. We show the results of the best fit to the experimental data in the Fig.\ref{fig:harmonics-fit}. The computed torque fields are: $|H_\mathrm{DL}| = 420 \frac{A}{m}$ and $|H_\mathrm{FL}| = 574 \frac{A}{m}$ at 5mA current, with the remaining parameters of the simulated structure taken from the Table\ref{tab:parameters}. As the experimental rotations may have a slight angular error, we emulate that in the simulation by allowing for the $\theta$ and $\phi$ (as per Fig.\ref{fig:coordinates}) angle to deviate from ideal up to $\approx 3^\circ$. 

\section{Parametric Analysis}
Finally, after introduction of the model and presentation of the best fits to the experimental results, we turn to the discussion of the parametric analysis. Specifically, we performed the analysis of $\mu_\textrm{0}M_\textrm{s}$, $K_\textrm{u}$, and $\theta$ by scanning each within $\pm 5\%$ margin respective to the value of best fit to the experimental data (Fig.\ref{fig:parameter-scan}). The deviations of the polar angle cause little effect for the second harmonics in the region of interest (the linear region in the low-field regime) but may contribute to large changes in the first harmonic at higher field magnitudes. From contrasting $\mu_\textrm{0}M_\textrm{s}$ and $K_\textrm{u}$ scans we see that the former has a much greater impact on both harmonic Hall voltage components than the latter -- specifically, much greater widening/stretching of the quadratic region in the first harmonic and the significant increase/decrease of the slope in the second harmonics in both arrangements. Furthermore, manipulating $\mu_\textrm{0}M_\textrm{s}$ has an inverse effect of that of $K_\textrm{u}$. Namely, an increase in $\mu_\textrm{0}M_\textrm{s}$ may potentially compensated by the adequate decrease of the $K_\textrm{u}$.

\begin{figure}[ht]
    \centering
    \includegraphics[width=\linewidth]{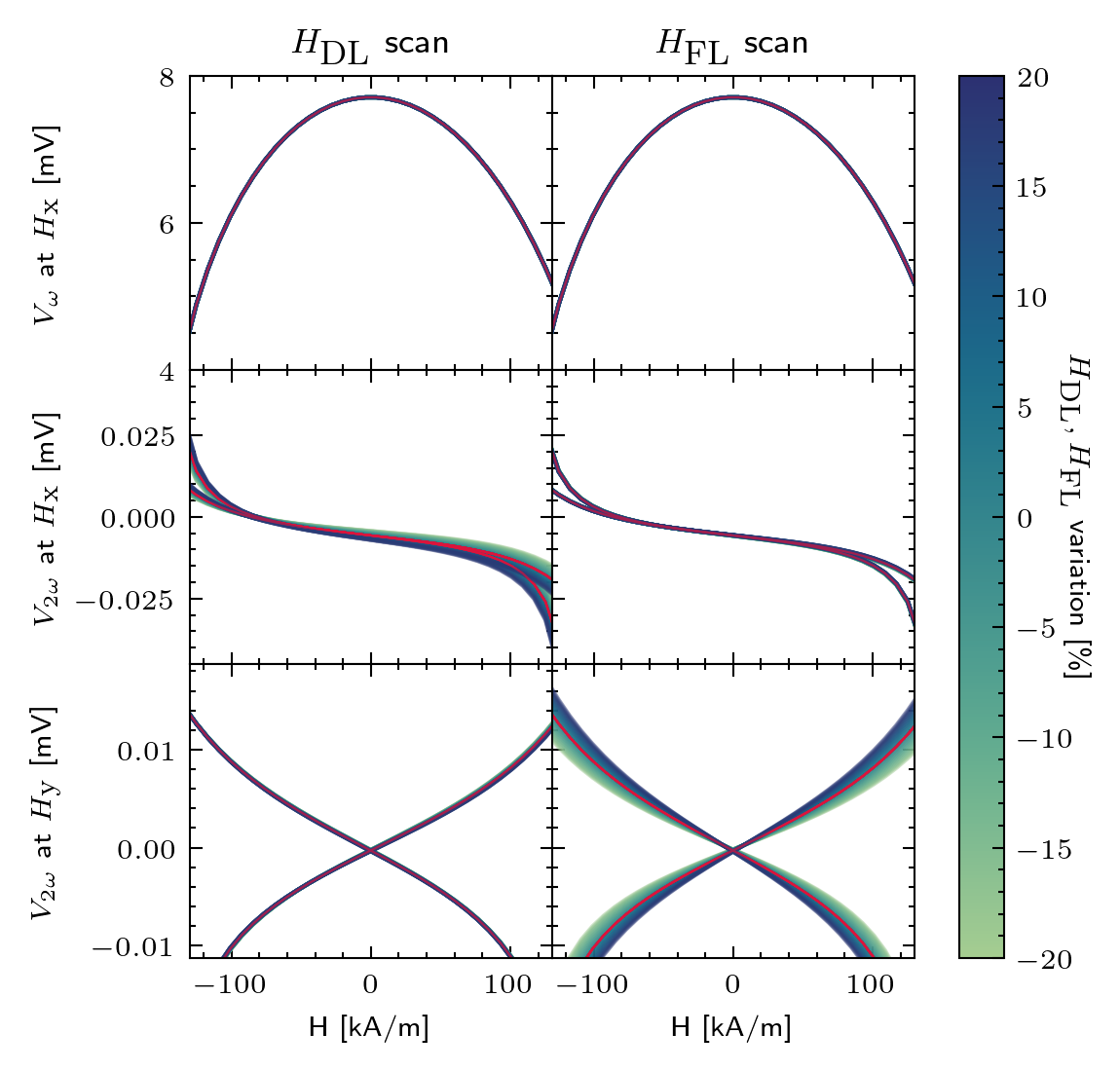}
    \caption{The torque variation at different field arrangements. Clearly, $H_\textrm{DL}$ influences only the 2nd harmonic at $H_\textrm{x}$ setting but has no strong effect when the sample is subjected to field at $H_\textrm{y}$. Conversely, $H_\textrm{FL}$ has a strong impact on the angle of line crossing in $H_\textrm{y}$ arrangement but causes little changes in $H_\textrm{x}$ setup. Neither of the torque contributions affects significantly the first harmonic components.}
    \label{fig:torque-scan}
\end{figure}

We then turn to the analysis of the behaviour of the FL- and DL-torques when their values are being varied while all other parameters stay constant (note that the magnitude of change is now $\pm 20\%$). The Fig.\ref{fig:torque-scan} depicts this attempt at evaluating the influence of torque modifications across different applied field arrangements. Firstly, we observe little to no change in the first harmonic response under either FL- or DL-torque variation. Altering the DL torque yields a significant deviation of the slope of the linear region in the longitudinal arrangement towards lower field regimes. However, manipulating with DL torque causes no visible alterations in the transverse arrangement. The situation flips when the FL torque is varied while DL torque is kept constant -- we notice a visible decrease in the slope values over the crossing linear regions in the $H_\textrm{y}$ arrangements, but no remarkable changes in the $H_\textrm{x}$ setting.
Together with the analysis of the Fig.\ref{fig:parameter-scan} we may posit that in our experiments, the first harmonic was entirely affected by the values of $\theta, \mu_\textrm{0}M_\textrm{s}$ or $K_\textrm{u}$. The inspection of the second harmonic components becomes more involved, as the linear regions may be compensated by manipulating with $\mu_\textrm{0}M_\textrm{s}, K_\textrm{u}$, but also the corresponding torque component and to a lesser degree also the $\theta$ angle. Fortunately, we can fix either $\mu_\textrm{0}M_\textrm{s}$ or $K_\textrm{u}$ by fitting to the dispersion relation first, thus reducing the initial problem of the second harmonic to tailoring exclusively $H_\textrm{FL}$ or $H_\textrm{DL}$ components.

\begin{figure}[ht]
    \centering
    \includegraphics[width=\linewidth]{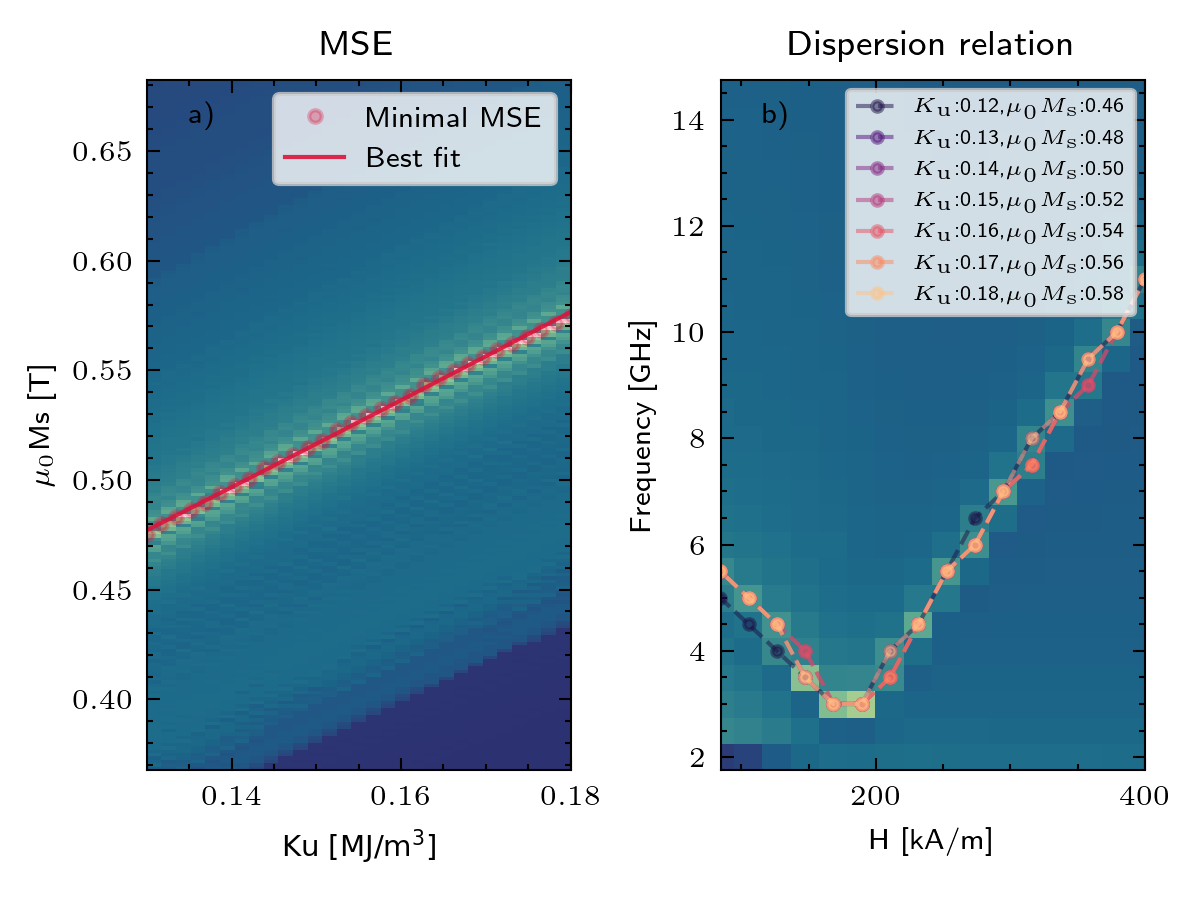}
    \caption{The map of the MSE error  respective to the SD-FMR generated with Table\ref{tab:parameters} (a). The brighter the colour the better minimum (smaller distance between two curves). (b) The resulting dispersion relation for a range of $K\textrm{u}$ (in [$\si{MJ/m^3}$]) parameters and respective $\mu_\textrm{0}M_\textrm{s}$ (in [$\si{T}$]) values computed from the fitted function in the right panel.}
    \label{fig:mse-param}
\end{figure}

In Fig.\ref{fig:mse-param} we plot the mean-squared error (MSE) between the parameters taken from the Table\ref{tab:parameters} and other simulations with parameters taken within some neighbourhood of the optimal ones. The colour indicates the magnitude of the MSE, with brighter regions corresponding to a lower MSE. From that figure we see that there is a line of minimal MSE for a range of $(\mu_\textrm{0}M_\textrm{s}, K_\textrm{u})$ pairs. For a small variation of the $K_\textrm{u}$, as is the case in the Fig.\ref{fig:mse-param}b), we may calculate the corresponding $\mu_\textrm{0}M_\textrm{s}$ values based on a linear model. Then, we overlay several SD-FMR lines computed based on those pairs to obtain Fig.\ref{fig:mse-param}a). We see that, within a good approximation, the $(\mu_\textrm{0}M_\textrm{s}, K_\textrm{u})$ pairs produce the same SD-FMR line, which supports the idea that there are several families of parameters that may be eligible for a fit. 
Hence, it is of primary importance to cross-check those values against the dispersion relation, as presented in Fig.\ref{fig:dynamics}. A good approximation of $\mu_\textrm{0}M_\textrm{s}$ from e.g. Vibrating Sample Magnetometer (VSM) measurements, helps in reducing the exhaustive search and narrow the range of feasible values of the $K_\textrm{u}$ parameter. To sum up, the parametric analysis shows specifically, that even a small variation of the saturation magnetisation may lead to significant over or underestimation of the SOT parameters. 

Combining the observations made regarding the Fig.\ref{fig:torque-scan} and Fig.\ref{fig:parameter-scan}, one may try to optimise for an optimal effective spin-Hall angle. Knowing how the change in $(\mu_\textrm{0}M_\textrm{s}, K_\textrm{u})$ pair affects the quadratic region of the first harmonic and the linear region of the second harmonic, it is possible to relate that change to the dependency from the Fig.\ref{fig:torque-scan} and thus conclude the effective impact on the spin-Hall angle, as computed in Eq.\ref{eq:hall-angle}.

\section{Conclusions}
To conclude, we demonstrated a stable, reproducible method for modelling a wide spectrum of static and dynamic experimental techniques using the macrospin numerical model. All parameters obtained from the experimental angular and field dependencies are consistent with  magnetoresistance dependencies, SD-FMR, and harmonics detection measurements. Furthermore, we performed scans of parameters such as $\mu_\textrm{0}M_\textrm{s}$, $K_\textrm{u}$ and $\theta$, each within $\pm 5\%$ to gain insight into the effect that those parameters have on the shape of harmonics. From that we show a strong dependency of both $V_\omega$ and $V_{2\omega}$ curves on $\mu_\textrm{0}M_\textrm{s}$, much more potent than that of $K_\textrm{u}$, which may have a significant bearing on the resultant Hall angle. However, here we presented only a small portion of the functionality offered by our model. Primarily, we focused on the determination of spin-torque effective fields $H_\mathrm{DL}$ and $H_\mathrm{FL}$, which has become an important problem in practical spintronics applications over recent years. CMTJ has more to offer in modelling multilayer spintronic devices and spintronics circuits, where several devices are coupled to each other via the mechanism of either electric or dipole coupling. Future work may also focus on automatic fitting parameters with Bayesian optimisation, significantly reducing the need of manual supervision of fit quality.

\section{Acknowledgements}
S.Ł., K.G., and T.S. acknowledge the National Science Centre, Poland, Grant No. Spinorbitronics UMO-2016/23/B/ST3/01430.
W.S. acknowledges the National Science Centre, Poland, Grant No. UMO-2015/17/D/ST3/00500. We would like to thank P. Ogrodnik and J.Chęciński for a fruitful discussion.

\appendix*
\section{Reformulating LLGS equation to the LL form}
\label{sec:appendix}
In this section, we outline the steps to obtain the numerically useful LL form of the LLGS equation wherein there there is no implicit $\frac{\textrm{d}\textbf{m}}{\textrm{dt}}$.
\begin{multline}
  \frac{\textrm{d}\textbf{m}}{\textrm{dt}} = -\gamma_0 \textbf{m} \times \textbf{H}_{\mathrm{eff}} + \alpha_\textrm{G} \textbf{m}\times \frac{\textrm{d}\textbf{m}}{\textrm{dt}} - \\ 
  \gamma_0|H_\textrm{FL}|\textbf{m} \times \mathbf{p}-\gamma_0|H_\textrm{DL}|\textbf{m} \times\textbf{m}\times \mathbf{p} 
 \label{eq:torque-llg}
\end{multline}
We follow \cite{camley_chapter_2012}. Firstly, applying $\mathbf{m} \times$ to the equation (\ref{eq:torque-llg}) yields: 
\begin{multline}
    \textbf{m} \times \frac{\textrm{d}\textbf{m}}{\textrm{dt}} = -\gamma_0 \textbf{m} \times\textbf{m} \times \textbf{H}_{\mathrm{eff}} + \alpha_\textrm{G} \textbf{m} \times \textbf{m}\times \frac{\textrm{d}\textbf{m}}{\textrm{dt}} - \\ \gamma_0|H_\textrm{FL}| \textbf{m} \times \textbf{m} \times \mathbf{p}- \gamma_0|H_\textrm{DL}|\textbf{m} \times \textbf{m}\times\textbf{m}\times \mathbf{p} 
    \label{eq:mcdm}
\end{multline}
After some simplification of (\ref{eq:mcdm}) we obtain:
\begin{multline}
    \textbf{m} \times \frac{\textrm{d}\textbf{m}}{\textrm{dt}} = -\gamma_0  \textbf{m} \times\textbf{m} \times \textbf{H}_{\mathrm{eff}} - \\
    \alpha_\textrm{G} \frac{\textrm{d}\textbf{m}}{\textrm{dt}}  -  \gamma_0|H_\textrm{FL}| \textbf{m} \times \textbf{m} \times \mathbf{p} + \gamma_0|H_\textrm{DL}|\textbf{m}\times \mathbf{p}
    \label{eq:RHS-dm}
\end{multline}
Substituting RHS of (\ref{eq:RHS-dm}) into (\ref{eq:torque-llg}) in lieu of $\textbf{m}\times \frac{\textrm{d}\textbf{m}}{\textrm{dt}}$ term leads to:
\begin{multline}
  \frac{\textrm{d}\textbf{m}}{\textrm{dt}} = -\gamma_0 \textbf{m} \times \textbf{H}_{\mathrm{eff}} + \alpha_\textrm{G} [-\gamma_0  \textbf{m} \times\textbf{m} \times \textbf{H}_{\mathrm{eff}} \\ - \alpha_\textrm{G} \frac{\textrm{d}\textbf{m}}{\textrm{dt}}  - \gamma_0|H_\textrm{FL}|\textbf{m} \times \textbf{m} \times \mathbf{p} + \gamma_0|H_\textrm{DL}|\textbf{m}\times \mathbf{p}] \\ -   
  \gamma_0|H_\textrm{FL}|\textbf{m} \times \mathbf{p} - 
    \gamma_0|H_\textrm{DL}|\textbf{m}\times\textbf{m}\times \mathbf{p}
    \nonumber
\end{multline}
Gathering all the $\frac{\textrm{d}\textbf{m}}{\textrm{dt}}$ terms produces:
\begin{multline}
     \frac{\textrm{d}\textbf{m}}{\textrm{dt}}(1 + \alpha_\textrm{G}^2) = -\gamma_0 \textbf{m} \times \textbf{H}_{\mathrm{eff}} - \alpha_\textrm{G}\gamma_0\textbf{m}\times\textbf{m}\times\textbf{H}_{\mathrm{eff}} \\ 
     - \gamma_0|H_\textrm{FL}|[\textbf{m} \times \mathbf{p} + \alpha_\textrm{G}\textbf{m} \times \textbf{m} \times \mathbf{p}] \\ 
     -  \gamma_0|H_\textrm{DL}|[\textbf{m}\times\textbf{m}\times \mathbf{p} - \alpha_\textrm{G}\textbf{m}\times \mathbf{p}] \nonumber
\end{multline}
Rearranging the torque terms gives:
\begin{widetext}
\begin{equation}
	\frac{\textrm{d}\textbf{m}}{\textrm{dt}} = \frac{-\gamma_0}{1 + \alpha_\textrm{G}^2}[\textbf{m} \times \textbf{H}_{\mathrm{eff}} + \alpha_\textrm{G}\textbf{m}\times\textbf{m}\times\textbf{H}_{\mathrm{eff}}]
     + \frac{-\gamma_0}{1 + \alpha_\textrm{G}^2}[|H_\textrm{FL}|[\textbf{m} \times \mathbf{p}  + \alpha_\textrm{G} \textbf{m} \times \textbf{m} \times \mathbf{p}] \nonumber\\
     + |H_\textrm{DL}|[\textbf{m}\times\textbf{m}\times \mathbf{p} - \alpha_\textrm{G}\textbf{m}\times \mathbf{p}]]
     \label{eq:no-mix}
\end{equation}
\end{widetext}
The last part of the Eq.\ref{eq:no-mix} can be rearranged to:
\begin{widetext}
\begin{multline}
\frac{\textrm{d}\textbf{m}}{\textrm{dt}} = \frac{-\gamma_0}{1 + \alpha_\textrm{G}^2}[\textbf{m} \times \textbf{H}_{\mathrm{eff}} + \alpha_\textrm{G}\textbf{m}\times\textbf{m}\times\textbf{H}_{\mathrm{eff}}] + \frac{-\gamma_0}{1 + \alpha_\textrm{G}^2}[\textbf{m} \times \mathbf{p}(|H_\textrm{FL}| - \alpha_\textrm{G}|H_\textrm{DL}|) + \textbf{m}\times\textbf{m}\times\mathbf{p}(|H_\textrm{DL}| + \alpha_\textrm{G}|H_\textrm{FL}|)]
\label{eq:numerical-ll}
\end{multline}
\end{widetext}

What becomes evident in this LL form of the LLG equation is the mixing of the torques with damping as the scaling factor. The field-like term, for instance, becomes $|H_\textrm{FL}| - \alpha|H_\textrm{DL}|$). We may neglect the second part of that term for small values of $|H_\textrm{FL}| \gg \alpha|H_\textrm{DL}|$. For numerical computation, we use the Eq.\ref{eq:numerical-ll}.

\bibliography{references}

\begin{thebibliography}{32}%
\makeatletter
\providecommand \@ifxundefined [1]{%
 \@ifx{#1\undefined}
}%
\providecommand \@ifnum [1]{%
 \ifnum #1\expandafter \@firstoftwo
 \else \expandafter \@secondoftwo
 \fi
}%
\providecommand \@ifx [1]{%
 \ifx #1\expandafter \@firstoftwo
 \else \expandafter \@secondoftwo
 \fi
}%
\providecommand \natexlab [1]{#1}%
\providecommand \enquote  [1]{``#1''}%
\providecommand \bibnamefont  [1]{#1}%
\providecommand \bibfnamefont [1]{#1}%
\providecommand \citenamefont [1]{#1}%
\providecommand \href@noop [0]{\@secondoftwo}%
\providecommand \href [0]{\begingroup \@sanitize@url \@href}%
\providecommand \@href[1]{\@@startlink{#1}\@@href}%
\providecommand \@@href[1]{\endgroup#1\@@endlink}%
\providecommand \@sanitize@url [0]{\catcode `\\12\catcode `\$12\catcode
  `\&12\catcode `\#12\catcode `\^12\catcode `\_12\catcode `\%12\relax}%
\providecommand \@@startlink[1]{}%
\providecommand \@@endlink[0]{}%
\providecommand \url  [0]{\begingroup\@sanitize@url \@url }%
\providecommand \@url [1]{\endgroup\@href {#1}{\urlprefix }}%
\providecommand \urlprefix  [0]{URL }%
\providecommand \Eprint [0]{\href }%
\providecommand \doibase [0]{https://doi.org/}%
\providecommand \selectlanguage [0]{\@gobble}%
\providecommand \bibinfo  [0]{\@secondoftwo}%
\providecommand \bibfield  [0]{\@secondoftwo}%
\providecommand \translation [1]{[#1]}%
\providecommand \BibitemOpen [0]{}%
\providecommand \bibitemStop [0]{}%
\providecommand \bibitemNoStop [0]{.\EOS\space}%
\providecommand \EOS [0]{\spacefactor3000\relax}%
\providecommand \BibitemShut  [1]{\csname bibitem#1\endcsname}%
\let\auto@bib@innerbib\@empty
\bibitem [{\citenamefont {Dieny}\ \emph {et~al.}(2020)\citenamefont {Dieny},
  \citenamefont {Prejbeanu}, \citenamefont {Garello}, \citenamefont
  {Gambardella}, \citenamefont {Freitas}, \citenamefont {Lehndorff},
  \citenamefont {Raberg}, \citenamefont {Ebels}, \citenamefont {Demokritov},
  \citenamefont {Akerman}, \citenamefont {Deac}, \citenamefont {Pirro},
  \citenamefont {Adelmann}, \citenamefont {Anane}, \citenamefont {Chumak},
  \citenamefont {Hirohata}, \citenamefont {Mangin}, \citenamefont {Valenzuela},
  \citenamefont {Onba{\c{s}}l{\i}}, \citenamefont {d'Aquino}, \citenamefont
  {Prenat}, \citenamefont {Finocchio}, \citenamefont {Lopez-Diaz},
  \citenamefont {Chantrell}, \citenamefont {Chubykalo-Fesenko},\ and\
  \citenamefont {Bortolotti}}]{Dieny2020}%
  \BibitemOpen
  \bibfield  {author} {\bibinfo {author} {\bibfnamefont {B.}~\bibnamefont
  {Dieny}}, \bibinfo {author} {\bibfnamefont {I.~L.}\ \bibnamefont
  {Prejbeanu}}, \bibinfo {author} {\bibfnamefont {K.}~\bibnamefont {Garello}},
  \bibinfo {author} {\bibfnamefont {P.}~\bibnamefont {Gambardella}}, \bibinfo
  {author} {\bibfnamefont {P.}~\bibnamefont {Freitas}}, \bibinfo {author}
  {\bibfnamefont {R.}~\bibnamefont {Lehndorff}}, \bibinfo {author}
  {\bibfnamefont {W.}~\bibnamefont {Raberg}}, \bibinfo {author} {\bibfnamefont
  {U.}~\bibnamefont {Ebels}}, \bibinfo {author} {\bibfnamefont {S.~O.}\
  \bibnamefont {Demokritov}}, \bibinfo {author} {\bibfnamefont
  {J.}~\bibnamefont {Akerman}}, \bibinfo {author} {\bibfnamefont
  {A.}~\bibnamefont {Deac}}, \bibinfo {author} {\bibfnamefont {P.}~\bibnamefont
  {Pirro}}, \bibinfo {author} {\bibfnamefont {C.}~\bibnamefont {Adelmann}},
  \bibinfo {author} {\bibfnamefont {A.}~\bibnamefont {Anane}}, \bibinfo
  {author} {\bibfnamefont {A.~V.}\ \bibnamefont {Chumak}}, \bibinfo {author}
  {\bibfnamefont {A.}~\bibnamefont {Hirohata}}, \bibinfo {author}
  {\bibfnamefont {S.}~\bibnamefont {Mangin}}, \bibinfo {author} {\bibfnamefont
  {S.~O.}\ \bibnamefont {Valenzuela}}, \bibinfo {author} {\bibfnamefont
  {M.~C.}\ \bibnamefont {Onba{\c{s}}l{\i}}}, \bibinfo {author} {\bibfnamefont
  {M.}~\bibnamefont {d'Aquino}}, \bibinfo {author} {\bibfnamefont
  {G.}~\bibnamefont {Prenat}}, \bibinfo {author} {\bibfnamefont
  {G.}~\bibnamefont {Finocchio}}, \bibinfo {author} {\bibfnamefont
  {L.}~\bibnamefont {Lopez-Diaz}}, \bibinfo {author} {\bibfnamefont
  {R.}~\bibnamefont {Chantrell}}, \bibinfo {author} {\bibfnamefont
  {O.}~\bibnamefont {Chubykalo-Fesenko}},\ and\ \bibinfo {author}
  {\bibfnamefont {P.}~\bibnamefont {Bortolotti}},\ }\bibfield  {title}
  {\bibinfo {title} {Opportunities and challenges for spintronics in the
  microelectronics industry},\ }\href
  {https://doi.org/10.1038/s41928-020-0461-5} {\bibfield  {journal} {\bibinfo
  {journal} {Nature Electronics}\ }\textbf {\bibinfo {volume} {3}},\ \bibinfo
  {pages} {446} (\bibinfo {year} {2020})}\BibitemShut {NoStop}%
\bibitem [{\citenamefont {Bhatti}\ \emph {et~al.}(2017)\citenamefont {Bhatti},
  \citenamefont {Sbiaa}, \citenamefont {Hirohata}, \citenamefont {Ohno},
  \citenamefont {Fukami},\ and\ \citenamefont
  {Piramanayagam}}]{bhatti2017spintronics}%
  \BibitemOpen
  \bibfield  {author} {\bibinfo {author} {\bibfnamefont {S.}~\bibnamefont
  {Bhatti}}, \bibinfo {author} {\bibfnamefont {R.}~\bibnamefont {Sbiaa}},
  \bibinfo {author} {\bibfnamefont {A.}~\bibnamefont {Hirohata}}, \bibinfo
  {author} {\bibfnamefont {H.}~\bibnamefont {Ohno}}, \bibinfo {author}
  {\bibfnamefont {S.}~\bibnamefont {Fukami}},\ and\ \bibinfo {author}
  {\bibfnamefont {S.}~\bibnamefont {Piramanayagam}},\ }\bibfield  {title}
  {\bibinfo {title} {Spintronics based random access memory: a review},\
  }\href@noop {} {\bibfield  {journal} {\bibinfo  {journal} {Materials Today}\
  }\textbf {\bibinfo {volume} {20}},\ \bibinfo {pages} {530} (\bibinfo {year}
  {2017})}\BibitemShut {NoStop}%
\bibitem [{\citenamefont {Ikegawa}\ \emph {et~al.}(2020)\citenamefont
  {Ikegawa}, \citenamefont {Mancoff}, \citenamefont {Janesky},\ and\
  \citenamefont {Aggarwal}}]{ikegawa2020magnetoresistive}%
  \BibitemOpen
  \bibfield  {author} {\bibinfo {author} {\bibfnamefont {S.}~\bibnamefont
  {Ikegawa}}, \bibinfo {author} {\bibfnamefont {F.~B.}\ \bibnamefont
  {Mancoff}}, \bibinfo {author} {\bibfnamefont {J.}~\bibnamefont {Janesky}},\
  and\ \bibinfo {author} {\bibfnamefont {S.}~\bibnamefont {Aggarwal}},\
  }\bibfield  {title} {\bibinfo {title} {Magnetoresistive random access memory:
  Present and future},\ }\href@noop {} {\bibfield  {journal} {\bibinfo
  {journal} {IEEE Transactions on Electron Devices}\ }\textbf {\bibinfo
  {volume} {67}},\ \bibinfo {pages} {1407} (\bibinfo {year}
  {2020})}\BibitemShut {NoStop}%
\bibitem [{\citenamefont {Hirohata}\ \emph {et~al.}(2020)\citenamefont
  {Hirohata}, \citenamefont {Yamada}, \citenamefont {Nakatani}, \citenamefont
  {Prejbeanu}, \citenamefont {Di{\'e}ny}, \citenamefont {Pirro},\ and\
  \citenamefont {Hillebrands}}]{hirohata2020review}%
  \BibitemOpen
  \bibfield  {author} {\bibinfo {author} {\bibfnamefont {A.}~\bibnamefont
  {Hirohata}}, \bibinfo {author} {\bibfnamefont {K.}~\bibnamefont {Yamada}},
  \bibinfo {author} {\bibfnamefont {Y.}~\bibnamefont {Nakatani}}, \bibinfo
  {author} {\bibfnamefont {I.-L.}\ \bibnamefont {Prejbeanu}}, \bibinfo {author}
  {\bibfnamefont {B.}~\bibnamefont {Di{\'e}ny}}, \bibinfo {author}
  {\bibfnamefont {P.}~\bibnamefont {Pirro}},\ and\ \bibinfo {author}
  {\bibfnamefont {B.}~\bibnamefont {Hillebrands}},\ }\bibfield  {title}
  {\bibinfo {title} {Review on spintronics: Principles and device
  applications},\ }\href@noop {} {\bibfield  {journal} {\bibinfo  {journal}
  {Journal of Magnetism and Magnetic Materials}\ }\textbf {\bibinfo {volume}
  {509}},\ \bibinfo {pages} {166711} (\bibinfo {year} {2020})}\BibitemShut
  {NoStop}%
\bibitem [{\citenamefont {Manipatruni}\ \emph {et~al.}(2019)\citenamefont
  {Manipatruni}, \citenamefont {Nikonov}, \citenamefont {Lin}, \citenamefont
  {Gosavi}, \citenamefont {Liu}, \citenamefont {Prasad}, \citenamefont {Huang},
  \citenamefont {Bonturim}, \citenamefont {Ramesh},\ and\ \citenamefont
  {Young}}]{Manipatruni2019}%
  \BibitemOpen
  \bibfield  {author} {\bibinfo {author} {\bibfnamefont {S.}~\bibnamefont
  {Manipatruni}}, \bibinfo {author} {\bibfnamefont {D.~E.}\ \bibnamefont
  {Nikonov}}, \bibinfo {author} {\bibfnamefont {C.-C.}\ \bibnamefont {Lin}},
  \bibinfo {author} {\bibfnamefont {T.~A.}\ \bibnamefont {Gosavi}}, \bibinfo
  {author} {\bibfnamefont {H.}~\bibnamefont {Liu}}, \bibinfo {author}
  {\bibfnamefont {B.}~\bibnamefont {Prasad}}, \bibinfo {author} {\bibfnamefont
  {Y.-L.}\ \bibnamefont {Huang}}, \bibinfo {author} {\bibfnamefont
  {E.}~\bibnamefont {Bonturim}}, \bibinfo {author} {\bibfnamefont
  {R.}~\bibnamefont {Ramesh}},\ and\ \bibinfo {author} {\bibfnamefont {I.~A.}\
  \bibnamefont {Young}},\ }\bibfield  {title} {\bibinfo {title} {Scalable
  energy-efficient magnetoelectric spin--orbit logic},\ }\href
  {https://doi.org/10.1038/s41586-018-0770-2} {\bibfield  {journal} {\bibinfo
  {journal} {Nature}\ }\textbf {\bibinfo {volume} {565}},\ \bibinfo {pages}
  {35} (\bibinfo {year} {2019})}\BibitemShut {NoStop}%
\bibitem [{\citenamefont {Ralph}\ and\ \citenamefont
  {Stiles}(2008)}]{ralph_spin_2008}%
  \BibitemOpen
  \bibfield  {author} {\bibinfo {author} {\bibfnamefont {D.}~\bibnamefont
  {Ralph}}\ and\ \bibinfo {author} {\bibfnamefont {M.}~\bibnamefont {Stiles}},\
  }\bibfield  {title} {\bibinfo {title} {Spin transfer torques},\ }\href
  {https://doi.org/10.1016/j.jmmm.2007.12.019} {\bibfield  {journal} {\bibinfo
  {journal} {Journal of Magnetism and Magnetic Materials}\ }\textbf {\bibinfo
  {volume} {320}},\ \bibinfo {pages} {1190} (\bibinfo {year}
  {2008})}\BibitemShut {NoStop}%
\bibitem [{\citenamefont {Brataas}\ \emph {et~al.}(2012)\citenamefont
  {Brataas}, \citenamefont {Kent},\ and\ \citenamefont
  {Ohno}}]{brataas2012current}%
  \BibitemOpen
  \bibfield  {author} {\bibinfo {author} {\bibfnamefont {A.}~\bibnamefont
  {Brataas}}, \bibinfo {author} {\bibfnamefont {A.~D.}\ \bibnamefont {Kent}},\
  and\ \bibinfo {author} {\bibfnamefont {H.}~\bibnamefont {Ohno}},\ }\bibfield
  {title} {\bibinfo {title} {Current-induced torques in magnetic materials},\
  }\href@noop {} {\bibfield  {journal} {\bibinfo  {journal} {Nature materials}\
  }\textbf {\bibinfo {volume} {11}},\ \bibinfo {pages} {372} (\bibinfo {year}
  {2012})}\BibitemShut {NoStop}%
\bibitem [{\citenamefont {Manchon}\ \emph {et~al.}(2019)\citenamefont
  {Manchon}, \citenamefont {{\v{Z}}elezn{\`y}}, \citenamefont {Miron},
  \citenamefont {Jungwirth}, \citenamefont {Sinova}, \citenamefont {Thiaville},
  \citenamefont {Garello},\ and\ \citenamefont
  {Gambardella}}]{manchon2019current}%
  \BibitemOpen
  \bibfield  {author} {\bibinfo {author} {\bibfnamefont {A.}~\bibnamefont
  {Manchon}}, \bibinfo {author} {\bibfnamefont {J.}~\bibnamefont
  {{\v{Z}}elezn{\`y}}}, \bibinfo {author} {\bibfnamefont {I.~M.}\ \bibnamefont
  {Miron}}, \bibinfo {author} {\bibfnamefont {T.}~\bibnamefont {Jungwirth}},
  \bibinfo {author} {\bibfnamefont {J.}~\bibnamefont {Sinova}}, \bibinfo
  {author} {\bibfnamefont {A.}~\bibnamefont {Thiaville}}, \bibinfo {author}
  {\bibfnamefont {K.}~\bibnamefont {Garello}},\ and\ \bibinfo {author}
  {\bibfnamefont {P.}~\bibnamefont {Gambardella}},\ }\bibfield  {title}
  {\bibinfo {title} {Current-induced spin-orbit torques in ferromagnetic and
  antiferromagnetic systems},\ }\href@noop {} {\bibfield  {journal} {\bibinfo
  {journal} {Reviews of Modern Physics}\ }\textbf {\bibinfo {volume} {91}},\
  \bibinfo {pages} {035004} (\bibinfo {year} {2019})}\BibitemShut {NoStop}%
\bibitem [{\citenamefont {Zhang}\ \emph {et~al.}(2021)\citenamefont {Zhang},
  \citenamefont {Takeuchi}, \citenamefont {Fukami},\ and\ \citenamefont
  {Ohno}}]{zhang2021field}%
  \BibitemOpen
  \bibfield  {author} {\bibinfo {author} {\bibfnamefont {C.}~\bibnamefont
  {Zhang}}, \bibinfo {author} {\bibfnamefont {Y.}~\bibnamefont {Takeuchi}},
  \bibinfo {author} {\bibfnamefont {S.}~\bibnamefont {Fukami}},\ and\ \bibinfo
  {author} {\bibfnamefont {H.}~\bibnamefont {Ohno}},\ }\bibfield  {title}
  {\bibinfo {title} {Field-free and sub-ns magnetization switching of magnetic
  tunnel junctions by combining spin-transfer torque and spin--orbit torque},\
  }\href@noop {} {\bibfield  {journal} {\bibinfo  {journal} {Applied Physics
  Letters}\ }\textbf {\bibinfo {volume} {118}},\ \bibinfo {pages} {092406}
  (\bibinfo {year} {2021})}\BibitemShut {NoStop}%
\bibitem [{\citenamefont {Chen}\ \emph {et~al.}(2020)\citenamefont {Chen},
  \citenamefont {Pan}, \citenamefont {Wang}, \citenamefont {Qiu}, \citenamefont
  {Lin}, \citenamefont {Liu}, \citenamefont {Li}, \citenamefont {Han},
  \citenamefont {Shi}, \citenamefont {Ando} \emph
  {et~al.}}]{chen2020manipulation}%
  \BibitemOpen
  \bibfield  {author} {\bibinfo {author} {\bibfnamefont {Z.}~\bibnamefont
  {Chen}}, \bibinfo {author} {\bibfnamefont {C.}~\bibnamefont {Pan}}, \bibinfo
  {author} {\bibfnamefont {N.}~\bibnamefont {Wang}}, \bibinfo {author}
  {\bibfnamefont {M.}~\bibnamefont {Qiu}}, \bibinfo {author} {\bibfnamefont
  {T.}~\bibnamefont {Lin}}, \bibinfo {author} {\bibfnamefont {J.}~\bibnamefont
  {Liu}}, \bibinfo {author} {\bibfnamefont {S.}~\bibnamefont {Li}}, \bibinfo
  {author} {\bibfnamefont {P.}~\bibnamefont {Han}}, \bibinfo {author}
  {\bibfnamefont {J.}~\bibnamefont {Shi}}, \bibinfo {author} {\bibfnamefont
  {K.}~\bibnamefont {Ando}}, \emph {et~al.},\ }\bibfield  {title} {\bibinfo
  {title} {Manipulation of perpendicular exchange bias and spin-orbit torques
  via mgo in pt/co/mgo films},\ }\href@noop {} {\bibfield  {journal} {\bibinfo
  {journal} {Journal of Magnetism and Magnetic Materials}\ }\textbf {\bibinfo
  {volume} {507}},\ \bibinfo {pages} {166822} (\bibinfo {year}
  {2020})}\BibitemShut {NoStop}%
\bibitem [{\citenamefont {Liu}\ \emph {et~al.}(2012)\citenamefont {Liu},
  \citenamefont {Pai}, \citenamefont {Li}, \citenamefont {Tseng}, \citenamefont
  {Ralph},\ and\ \citenamefont {Buhrman}}]{liu2012spin}%
  \BibitemOpen
  \bibfield  {author} {\bibinfo {author} {\bibfnamefont {L.}~\bibnamefont
  {Liu}}, \bibinfo {author} {\bibfnamefont {C.-F.}\ \bibnamefont {Pai}},
  \bibinfo {author} {\bibfnamefont {Y.}~\bibnamefont {Li}}, \bibinfo {author}
  {\bibfnamefont {H.}~\bibnamefont {Tseng}}, \bibinfo {author} {\bibfnamefont
  {D.}~\bibnamefont {Ralph}},\ and\ \bibinfo {author} {\bibfnamefont
  {R.}~\bibnamefont {Buhrman}},\ }\bibfield  {title} {\bibinfo {title}
  {Spin-torque switching with the giant spin hall effect of tantalum},\
  }\href@noop {} {\bibfield  {journal} {\bibinfo  {journal} {Science}\ }\textbf
  {\bibinfo {volume} {336}},\ \bibinfo {pages} {555} (\bibinfo {year}
  {2012})}\BibitemShut {NoStop}%
\bibitem [{\citenamefont {Song}\ \emph {et~al.}(2021)\citenamefont {Song},
  \citenamefont {Zhang}, \citenamefont {Liao}, \citenamefont {Zhou},
  \citenamefont {Zhou}, \citenamefont {Chen}, \citenamefont {You},
  \citenamefont {Chen},\ and\ \citenamefont {Pan}}]{song2021spin}%
  \BibitemOpen
  \bibfield  {author} {\bibinfo {author} {\bibfnamefont {C.}~\bibnamefont
  {Song}}, \bibinfo {author} {\bibfnamefont {R.}~\bibnamefont {Zhang}},
  \bibinfo {author} {\bibfnamefont {L.}~\bibnamefont {Liao}}, \bibinfo {author}
  {\bibfnamefont {Y.}~\bibnamefont {Zhou}}, \bibinfo {author} {\bibfnamefont
  {X.}~\bibnamefont {Zhou}}, \bibinfo {author} {\bibfnamefont {R.}~\bibnamefont
  {Chen}}, \bibinfo {author} {\bibfnamefont {Y.}~\bibnamefont {You}}, \bibinfo
  {author} {\bibfnamefont {X.}~\bibnamefont {Chen}},\ and\ \bibinfo {author}
  {\bibfnamefont {F.}~\bibnamefont {Pan}},\ }\bibfield  {title} {\bibinfo
  {title} {Spin-orbit torques: Materials, mechanisms, performances, and
  potential applications},\ }\href@noop {} {\bibfield  {journal} {\bibinfo
  {journal} {Progress in Materials Science}\ }\textbf {\bibinfo {volume}
  {118}},\ \bibinfo {pages} {100761} (\bibinfo {year} {2021})}\BibitemShut
  {NoStop}%
\bibitem [{\citenamefont {Skowro{\'n}ski}\ \emph {et~al.}(2019)\citenamefont
  {Skowro{\'n}ski}, \citenamefont {Karwacki}, \citenamefont {Zi{\k{e}}tek},
  \citenamefont {Kanak}, \citenamefont {{\L}azarski}, \citenamefont {Grochot},
  \citenamefont {Stobiecki}, \citenamefont {Ku{\'s}wik}, \citenamefont
  {Stobiecki},\ and\ \citenamefont {Barna{\'s}}}]{skowronski2019determination}%
  \BibitemOpen
  \bibfield  {author} {\bibinfo {author} {\bibfnamefont {W.}~\bibnamefont
  {Skowro{\'n}ski}}, \bibinfo {author} {\bibfnamefont {{\L}.}~\bibnamefont
  {Karwacki}}, \bibinfo {author} {\bibfnamefont {S.}~\bibnamefont
  {Zi{\k{e}}tek}}, \bibinfo {author} {\bibfnamefont {J.}~\bibnamefont {Kanak}},
  \bibinfo {author} {\bibfnamefont {S.}~\bibnamefont {{\L}azarski}}, \bibinfo
  {author} {\bibfnamefont {K.}~\bibnamefont {Grochot}}, \bibinfo {author}
  {\bibfnamefont {T.}~\bibnamefont {Stobiecki}}, \bibinfo {author}
  {\bibfnamefont {P.}~\bibnamefont {Ku{\'s}wik}}, \bibinfo {author}
  {\bibfnamefont {F.}~\bibnamefont {Stobiecki}},\ and\ \bibinfo {author}
  {\bibfnamefont {J.}~\bibnamefont {Barna{\'s}}},\ }\bibfield  {title}
  {\bibinfo {title} {Determination of spin hall angle in heavy-metal/co- fe-
  b-based heterostructures with interfacial spin-orbit fields},\ }\href@noop {}
  {\bibfield  {journal} {\bibinfo  {journal} {Physical Review Applied}\
  }\textbf {\bibinfo {volume} {11}},\ \bibinfo {pages} {024039} (\bibinfo
  {year} {2019})}\BibitemShut {NoStop}%
\bibitem [{\citenamefont {Ogrodnik}\ \emph {et~al.}(2021)\citenamefont
  {Ogrodnik}, \citenamefont {Grochot}, \citenamefont {Karwacki}, \citenamefont
  {Kanak}, \citenamefont {Prokop}, \citenamefont {Ch{\k{e}}ci{\'n}ski},
  \citenamefont {Skowro{\'n}ski}, \citenamefont {Zi{\k{e}}tek},\ and\
  \citenamefont {Stobiecki}}]{ogrodnik2021study}%
  \BibitemOpen
  \bibfield  {author} {\bibinfo {author} {\bibfnamefont {P.}~\bibnamefont
  {Ogrodnik}}, \bibinfo {author} {\bibfnamefont {K.}~\bibnamefont {Grochot}},
  \bibinfo {author} {\bibfnamefont {{\L}.}~\bibnamefont {Karwacki}}, \bibinfo
  {author} {\bibfnamefont {J.}~\bibnamefont {Kanak}}, \bibinfo {author}
  {\bibfnamefont {M.}~\bibnamefont {Prokop}}, \bibinfo {author} {\bibfnamefont
  {J.}~\bibnamefont {Ch{\k{e}}ci{\'n}ski}}, \bibinfo {author} {\bibfnamefont
  {W.}~\bibnamefont {Skowro{\'n}ski}}, \bibinfo {author} {\bibfnamefont
  {S.}~\bibnamefont {Zi{\k{e}}tek}},\ and\ \bibinfo {author} {\bibfnamefont
  {T.}~\bibnamefont {Stobiecki}},\ }\bibfield  {title} {\bibinfo {title} {Study
  of spin-orbit interactions and interlayer ferromagnetic coupling in co/pt/co
  trilayers in wide range of heavy metal thickness},\ }\href@noop {} {\bibfield
   {journal} {\bibinfo  {journal} {ACS Applied Materials and Interfaces}\
  }\textbf {\bibinfo {volume} {13}},\ \bibinfo {pages} {47019} (\bibinfo {year}
  {2021})}\BibitemShut {NoStop}%
\bibitem [{\citenamefont {Grimaldi}\ \emph {et~al.}(2020)\citenamefont
  {Grimaldi}, \citenamefont {Krizakova}, \citenamefont {Sala}, \citenamefont
  {Yasin}, \citenamefont {Couet}, \citenamefont {Kar}, \citenamefont
  {Garello},\ and\ \citenamefont {Gambardella}}]{grimaldi2020single}%
  \BibitemOpen
  \bibfield  {author} {\bibinfo {author} {\bibfnamefont {E.}~\bibnamefont
  {Grimaldi}}, \bibinfo {author} {\bibfnamefont {V.}~\bibnamefont {Krizakova}},
  \bibinfo {author} {\bibfnamefont {G.}~\bibnamefont {Sala}}, \bibinfo {author}
  {\bibfnamefont {F.}~\bibnamefont {Yasin}}, \bibinfo {author} {\bibfnamefont
  {S.}~\bibnamefont {Couet}}, \bibinfo {author} {\bibfnamefont {G.~S.}\
  \bibnamefont {Kar}}, \bibinfo {author} {\bibfnamefont {K.}~\bibnamefont
  {Garello}},\ and\ \bibinfo {author} {\bibfnamefont {P.}~\bibnamefont
  {Gambardella}},\ }\bibfield  {title} {\bibinfo {title} {Single-shot dynamics
  of spin--orbit torque and spin transfer torque switching in three-terminal
  magnetic tunnel junctions},\ }\href@noop {} {\bibfield  {journal} {\bibinfo
  {journal} {Nature nanotechnology}\ }\textbf {\bibinfo {volume} {15}},\
  \bibinfo {pages} {111} (\bibinfo {year} {2020})}\BibitemShut {NoStop}%
\bibitem [{\citenamefont {Zhou}\ \emph {et~al.}(2020)\citenamefont {Zhou},
  \citenamefont {Wang}, \citenamefont {Li}, \citenamefont {Wang}, \citenamefont
  {Liu}, \citenamefont {Wu},\ and\ \citenamefont {Zhao}}]{zhou2020design}%
  \BibitemOpen
  \bibfield  {author} {\bibinfo {author} {\bibfnamefont {H.}~\bibnamefont
  {Zhou}}, \bibinfo {author} {\bibfnamefont {C.}~\bibnamefont {Wang}}, \bibinfo
  {author} {\bibfnamefont {Z.}~\bibnamefont {Li}}, \bibinfo {author}
  {\bibfnamefont {Z.}~\bibnamefont {Wang}}, \bibinfo {author} {\bibfnamefont
  {T.}~\bibnamefont {Liu}}, \bibinfo {author} {\bibfnamefont {B.}~\bibnamefont
  {Wu}},\ and\ \bibinfo {author} {\bibfnamefont {W.}~\bibnamefont {Zhao}},\
  }\bibfield  {title} {\bibinfo {title} {Design of an erasable spintronics
  memory based on current-path-dependent field-free spin orbit torque},\
  }\href@noop {} {\bibfield  {journal} {\bibinfo  {journal} {AIP Advances}\
  }\textbf {\bibinfo {volume} {10}},\ \bibinfo {pages} {015317} (\bibinfo
  {year} {2020})}\BibitemShut {NoStop}%
\bibitem [{\citenamefont {Liu}\ \emph {et~al.}(2011)\citenamefont {Liu},
  \citenamefont {Moriyama}, \citenamefont {Ralph},\ and\ \citenamefont
  {Buhrman}}]{liu2011spin}%
  \BibitemOpen
  \bibfield  {author} {\bibinfo {author} {\bibfnamefont {L.}~\bibnamefont
  {Liu}}, \bibinfo {author} {\bibfnamefont {T.}~\bibnamefont {Moriyama}},
  \bibinfo {author} {\bibfnamefont {D.}~\bibnamefont {Ralph}},\ and\ \bibinfo
  {author} {\bibfnamefont {R.}~\bibnamefont {Buhrman}},\ }\bibfield  {title}
  {\bibinfo {title} {Spin-torque ferromagnetic resonance induced by the spin
  hall effect},\ }\href@noop {} {\bibfield  {journal} {\bibinfo  {journal}
  {Physical review letters}\ }\textbf {\bibinfo {volume} {106}},\ \bibinfo
  {pages} {036601} (\bibinfo {year} {2011})}\BibitemShut {NoStop}%
\bibitem [{\citenamefont {Hao}\ and\ \citenamefont
  {Xiao}(2015)}]{hao2015giant}%
  \BibitemOpen
  \bibfield  {author} {\bibinfo {author} {\bibfnamefont {Q.}~\bibnamefont
  {Hao}}\ and\ \bibinfo {author} {\bibfnamefont {G.}~\bibnamefont {Xiao}},\
  }\bibfield  {title} {\bibinfo {title} {Giant spin hall effect and switching
  induced by spin-transfer torque in a w/co 40 fe 40 b 20/mgo structure with
  perpendicular magnetic anisotropy},\ }\href@noop {} {\bibfield  {journal}
  {\bibinfo  {journal} {Physical Review Applied}\ }\textbf {\bibinfo {volume}
  {3}},\ \bibinfo {pages} {034009} (\bibinfo {year} {2015})}\BibitemShut
  {NoStop}%
\bibitem [{\citenamefont {Hayashi}\ \emph {et~al.}()\citenamefont {Hayashi},
  \citenamefont {Kim}, \citenamefont {Yamanouchi},\ and\ \citenamefont
  {Ohno}}]{hayashi_quantitative_2014}%
  \BibitemOpen
  \bibfield  {author} {\bibinfo {author} {\bibfnamefont {M.}~\bibnamefont
  {Hayashi}}, \bibinfo {author} {\bibfnamefont {J.}~\bibnamefont {Kim}},
  \bibinfo {author} {\bibfnamefont {M.}~\bibnamefont {Yamanouchi}},\ and\
  \bibinfo {author} {\bibfnamefont {H.}~\bibnamefont {Ohno}},\ }\bibfield
  {title} {\bibinfo {title} {Quantitative characterization of the spin-orbit
  torque using harmonic hall voltage measurements},\ }\href
  {https://doi.org/10.1103/PhysRevB.89.144425} {\bibfield  {journal} {\bibinfo
  {journal} {Physical Review B}\ }\textbf {\bibinfo {volume} {89}},\ \bibinfo
  {pages} {144425}}\BibitemShut {NoStop}%
\bibitem [{\citenamefont {Kim}\ \emph {et~al.}(2013)\citenamefont {Kim},
  \citenamefont {Sinha}, \citenamefont {Hayashi}, \citenamefont {Yamanouchi},
  \citenamefont {Fukami}, \citenamefont {Suzuki}, \citenamefont {Mitani},\ and\
  \citenamefont {Ohno}}]{kim2013layer}%
  \BibitemOpen
  \bibfield  {author} {\bibinfo {author} {\bibfnamefont {J.}~\bibnamefont
  {Kim}}, \bibinfo {author} {\bibfnamefont {J.}~\bibnamefont {Sinha}}, \bibinfo
  {author} {\bibfnamefont {M.}~\bibnamefont {Hayashi}}, \bibinfo {author}
  {\bibfnamefont {M.}~\bibnamefont {Yamanouchi}}, \bibinfo {author}
  {\bibfnamefont {S.}~\bibnamefont {Fukami}}, \bibinfo {author} {\bibfnamefont
  {T.}~\bibnamefont {Suzuki}}, \bibinfo {author} {\bibfnamefont
  {S.}~\bibnamefont {Mitani}},\ and\ \bibinfo {author} {\bibfnamefont
  {H.}~\bibnamefont {Ohno}},\ }\bibfield  {title} {\bibinfo {title} {Layer
  thickness dependence of the current-induced effective field vector in ta|
  cofeb| mgo},\ }\href@noop {} {\bibfield  {journal} {\bibinfo  {journal}
  {Nature materials}\ }\textbf {\bibinfo {volume} {12}},\ \bibinfo {pages}
  {240} (\bibinfo {year} {2013})}\BibitemShut {NoStop}%
\bibitem [{\citenamefont {Nguyen}\ and\ \citenamefont
  {Pai}(2021)}]{nguyen_spinorbit_2021}%
  \BibitemOpen
  \bibfield  {author} {\bibinfo {author} {\bibfnamefont {M.-H.}\ \bibnamefont
  {Nguyen}}\ and\ \bibinfo {author} {\bibfnamefont {C.-F.}\ \bibnamefont
  {Pai}},\ }\bibfield  {title} {\bibinfo {title} {Spin–orbit torque
  characterization in a nutshell},\ }\href {https://doi.org/10.1063/5.0041123}
  {\bibfield  {journal} {\bibinfo  {journal} {APL Materials}\ }\textbf
  {\bibinfo {volume} {9}},\ \bibinfo {pages} {030902} (\bibinfo {year}
  {2021})}\BibitemShut {NoStop}%
\bibitem [{\citenamefont {Kim}\ \emph {et~al.}(2016)\citenamefont {Kim},
  \citenamefont {Sheng}, \citenamefont {Takahashi}, \citenamefont {Mitani},\
  and\ \citenamefont {Hayashi}}]{kim2016}%
  \BibitemOpen
  \bibfield  {author} {\bibinfo {author} {\bibfnamefont {J.}~\bibnamefont
  {Kim}}, \bibinfo {author} {\bibfnamefont {P.}~\bibnamefont {Sheng}}, \bibinfo
  {author} {\bibfnamefont {S.}~\bibnamefont {Takahashi}}, \bibinfo {author}
  {\bibfnamefont {S.}~\bibnamefont {Mitani}},\ and\ \bibinfo {author}
  {\bibfnamefont {M.}~\bibnamefont {Hayashi}},\ }\bibfield  {title} {\bibinfo
  {title} {Spin hall magnetoresistance in metallic bilayers},\ }\href
  {https://doi.org/10.1103/PhysRevLett.116.097201} {\bibfield  {journal}
  {\bibinfo  {journal} {Phys. Rev. Lett.}\ }\textbf {\bibinfo {volume} {116}},\
  \bibinfo {pages} {097201} (\bibinfo {year} {2016})}\BibitemShut {NoStop}%
\bibitem [{\citenamefont {Locatelli}\ \emph {et~al.}(2014)\citenamefont
  {Locatelli}, \citenamefont {Cros},\ and\ \citenamefont
  {Grollier}}]{Locatelli2014}%
  \BibitemOpen
  \bibfield  {author} {\bibinfo {author} {\bibfnamefont {N.}~\bibnamefont
  {Locatelli}}, \bibinfo {author} {\bibfnamefont {V.}~\bibnamefont {Cros}},\
  and\ \bibinfo {author} {\bibfnamefont {J.}~\bibnamefont {Grollier}},\
  }\bibfield  {title} {\bibinfo {title} {Spin-torque building blocks},\ }\href
  {https://doi.org/10.1038/nmat3823} {\bibfield  {journal} {\bibinfo  {journal}
  {Nature Materials}\ }\textbf {\bibinfo {volume} {13}},\ \bibinfo {pages} {11}
  (\bibinfo {year} {2014})}\BibitemShut {NoStop}%
\bibitem [{\citenamefont {Tulapurkar}\ \emph {et~al.}(2005)\citenamefont
  {Tulapurkar}, \citenamefont {Suzuki}, \citenamefont {Fukushima},
  \citenamefont {Kubota}, \citenamefont {Maehara}, \citenamefont {Tsunekawa},
  \citenamefont {Djayaprawira}, \citenamefont {Watanabe},\ and\ \citenamefont
  {Yuasa}}]{Tulapurkar2005}%
  \BibitemOpen
  \bibfield  {author} {\bibinfo {author} {\bibfnamefont {A.~A.}\ \bibnamefont
  {Tulapurkar}}, \bibinfo {author} {\bibfnamefont {Y.}~\bibnamefont {Suzuki}},
  \bibinfo {author} {\bibfnamefont {A.}~\bibnamefont {Fukushima}}, \bibinfo
  {author} {\bibfnamefont {H.}~\bibnamefont {Kubota}}, \bibinfo {author}
  {\bibfnamefont {H.}~\bibnamefont {Maehara}}, \bibinfo {author} {\bibfnamefont
  {K.}~\bibnamefont {Tsunekawa}}, \bibinfo {author} {\bibfnamefont {D.~D.}\
  \bibnamefont {Djayaprawira}}, \bibinfo {author} {\bibfnamefont
  {N.}~\bibnamefont {Watanabe}},\ and\ \bibinfo {author} {\bibfnamefont
  {S.}~\bibnamefont {Yuasa}},\ }\bibfield  {title} {\bibinfo {title}
  {Spin-torque diode effect in magnetic tunnel junctions},\ }\href
  {https://doi.org/10.1038/nature04207} {\bibfield  {journal} {\bibinfo
  {journal} {Nature}\ }\textbf {\bibinfo {volume} {438}},\ \bibinfo {pages}
  {339} (\bibinfo {year} {2005})}\BibitemShut {NoStop}%
\bibitem [{\citenamefont {Sankey}\ \emph {et~al.}(2008)\citenamefont {Sankey},
  \citenamefont {Cui}, \citenamefont {Sun}, \citenamefont {Slonczewski},
  \citenamefont {Buhrman},\ and\ \citenamefont
  {Ralph}}]{sankey2008measurement}%
  \BibitemOpen
  \bibfield  {author} {\bibinfo {author} {\bibfnamefont {J.~C.}\ \bibnamefont
  {Sankey}}, \bibinfo {author} {\bibfnamefont {Y.-T.}\ \bibnamefont {Cui}},
  \bibinfo {author} {\bibfnamefont {J.~Z.}\ \bibnamefont {Sun}}, \bibinfo
  {author} {\bibfnamefont {J.~C.}\ \bibnamefont {Slonczewski}}, \bibinfo
  {author} {\bibfnamefont {R.~A.}\ \bibnamefont {Buhrman}},\ and\ \bibinfo
  {author} {\bibfnamefont {D.~C.}\ \bibnamefont {Ralph}},\ }\bibfield  {title}
  {\bibinfo {title} {Measurement of the spin-transfer-torque vector in magnetic
  tunnel junctions},\ }\href@noop {} {\bibfield  {journal} {\bibinfo  {journal}
  {Nature Physics}\ }\textbf {\bibinfo {volume} {4}},\ \bibinfo {pages} {67}
  (\bibinfo {year} {2008})}\BibitemShut {NoStop}%
\bibitem [{\citenamefont {Zi{\k{e}}tek}\ \emph {et~al.}(2015)\citenamefont
  {Zi{\k{e}}tek}, \citenamefont {Ogrodnik}, \citenamefont {Frankowski},
  \citenamefont {Ch{\k{e}}ci{\'n}ski}, \citenamefont {Wi{\'s}niowski},
  \citenamefont {Skowro{\'n}ski}, \citenamefont {Wrona}, \citenamefont
  {Stobiecki}, \citenamefont {{\.Z}ywczak},\ and\ \citenamefont
  {Barna{\'s}}}]{ziketek2015rectification}%
  \BibitemOpen
  \bibfield  {author} {\bibinfo {author} {\bibfnamefont {S.}~\bibnamefont
  {Zi{\k{e}}tek}}, \bibinfo {author} {\bibfnamefont {P.}~\bibnamefont
  {Ogrodnik}}, \bibinfo {author} {\bibfnamefont {M.}~\bibnamefont
  {Frankowski}}, \bibinfo {author} {\bibfnamefont {J.}~\bibnamefont
  {Ch{\k{e}}ci{\'n}ski}}, \bibinfo {author} {\bibfnamefont {P.}~\bibnamefont
  {Wi{\'s}niowski}}, \bibinfo {author} {\bibfnamefont {W.}~\bibnamefont
  {Skowro{\'n}ski}}, \bibinfo {author} {\bibfnamefont {J.}~\bibnamefont
  {Wrona}}, \bibinfo {author} {\bibfnamefont {T.}~\bibnamefont {Stobiecki}},
  \bibinfo {author} {\bibfnamefont {A.}~\bibnamefont {{\.Z}ywczak}},\ and\
  \bibinfo {author} {\bibfnamefont {J.}~\bibnamefont {Barna{\'s}}},\ }\bibfield
   {title} {\bibinfo {title} {Rectification of radio-frequency current in a
  giant-magnetoresistance spin valve},\ }\href@noop {} {\bibfield  {journal}
  {\bibinfo  {journal} {Physical Review B}\ }\textbf {\bibinfo {volume} {91}},\
  \bibinfo {pages} {014430} (\bibinfo {year} {2015})}\BibitemShut {NoStop}%
\bibitem [{\citenamefont {Avci}\ \emph {et~al.}(2014)\citenamefont {Avci},
  \citenamefont {Garello}, \citenamefont {Gabureac}, \citenamefont {Ghosh},
  \citenamefont {Fuhrer}, \citenamefont {Alvarado},\ and\ \citenamefont
  {Gambardella}}]{avci2014}%
  \BibitemOpen
  \bibfield  {author} {\bibinfo {author} {\bibfnamefont {C.~O.}\ \bibnamefont
  {Avci}}, \bibinfo {author} {\bibfnamefont {K.}~\bibnamefont {Garello}},
  \bibinfo {author} {\bibfnamefont {M.}~\bibnamefont {Gabureac}}, \bibinfo
  {author} {\bibfnamefont {A.}~\bibnamefont {Ghosh}}, \bibinfo {author}
  {\bibfnamefont {A.}~\bibnamefont {Fuhrer}}, \bibinfo {author} {\bibfnamefont
  {S.~F.}\ \bibnamefont {Alvarado}},\ and\ \bibinfo {author} {\bibfnamefont
  {P.}~\bibnamefont {Gambardella}},\ }\bibfield  {title} {\bibinfo {title}
  {Interplay of spin-orbit torque and thermoelectric effects in
  ferromagnet/normal-metal bilayers},\ }\href
  {https://link.aps.org/doi/10.1103/PhysRevB.90.224427} {\bibfield  {journal}
  {\bibinfo  {journal} {Phys. Rev. B}\ }\textbf {\bibinfo {volume} {90}},\
  \bibinfo {pages} {224427} (\bibinfo {year} {2014})}\BibitemShut {NoStop}%
\bibitem [{loc(2011)}]{lockin}%
  \BibitemOpen
  \href {https://www.thinksrs.com/downloads/pdfs/manuals/SR830m.pdf} {\bibinfo
  {title} {Sr830 dsp lock-in amplifier manual}} (\bibinfo {year}
  {2011})\BibitemShut {NoStop}%
\bibitem [{\citenamefont {Cecot}\ \emph {et~al.}(2017)\citenamefont {Cecot},
  \citenamefont {Karwacki}, \citenamefont {Skowro{\'n}ski}, \citenamefont
  {Kanak}, \citenamefont {Wrona}, \citenamefont {{\.Z}ywczak}, \citenamefont
  {Yao}, \citenamefont {van Dijken}, \citenamefont {Barna{\'s}},\ and\
  \citenamefont {Stobiecki}}]{cecot2017influence}%
  \BibitemOpen
  \bibfield  {author} {\bibinfo {author} {\bibfnamefont {M.}~\bibnamefont
  {Cecot}}, \bibinfo {author} {\bibfnamefont {{\L}.}~\bibnamefont {Karwacki}},
  \bibinfo {author} {\bibfnamefont {W.}~\bibnamefont {Skowro{\'n}ski}},
  \bibinfo {author} {\bibfnamefont {J.}~\bibnamefont {Kanak}}, \bibinfo
  {author} {\bibfnamefont {J.}~\bibnamefont {Wrona}}, \bibinfo {author}
  {\bibfnamefont {A.}~\bibnamefont {{\.Z}ywczak}}, \bibinfo {author}
  {\bibfnamefont {L.}~\bibnamefont {Yao}}, \bibinfo {author} {\bibfnamefont
  {S.}~\bibnamefont {van Dijken}}, \bibinfo {author} {\bibfnamefont
  {J.}~\bibnamefont {Barna{\'s}}},\ and\ \bibinfo {author} {\bibfnamefont
  {T.}~\bibnamefont {Stobiecki}},\ }\bibfield  {title} {\bibinfo {title}
  {Influence of intermixing at the ta/cofeb interface on spin hall angle in
  ta/cofeb/mgo heterostructures},\ }\href@noop {} {\bibfield  {journal}
  {\bibinfo  {journal} {Scientific reports}\ }\textbf {\bibinfo {volume} {7}},\
  \bibinfo {pages} {1} (\bibinfo {year} {2017})}\BibitemShut {NoStop}%
\bibitem [{\citenamefont {\L{}azarski}\ \emph {et~al.}(2021)\citenamefont
  {\L{}azarski}, \citenamefont {Skowro\ifmmode~\acute{n}\else \'{n}\fi{}ski},
  \citenamefont {Grochot}, \citenamefont {Powro\ifmmode~\acute{z}\else
  \'{z}\fi{}nik}, \citenamefont {Kanak}, \citenamefont {Schmidt},\ and\
  \citenamefont {Stobiecki}}]{lazarski2021}%
  \BibitemOpen
  \bibfield  {author} {\bibinfo {author} {\bibfnamefont {S.}~\bibnamefont
  {\L{}azarski}}, \bibinfo {author} {\bibfnamefont {W.}~\bibnamefont
  {Skowro\ifmmode~\acute{n}\else \'{n}\fi{}ski}}, \bibinfo {author}
  {\bibfnamefont {K.}~\bibnamefont {Grochot}}, \bibinfo {author} {\bibfnamefont
  {W.}~\bibnamefont {Powro\ifmmode~\acute{z}\else \'{z}\fi{}nik}}, \bibinfo
  {author} {\bibfnamefont {J.}~\bibnamefont {Kanak}}, \bibinfo {author}
  {\bibfnamefont {M.}~\bibnamefont {Schmidt}},\ and\ \bibinfo {author}
  {\bibfnamefont {T.}~\bibnamefont {Stobiecki}},\ }\bibfield  {title} {\bibinfo
  {title} {Spin-orbit torque induced magnetization dynamics and switching in a
  cofeb/ta/cofeb system with mixed magnetic anisotropy},\ }\href
  {https://doi.org/10.1103/PhysRevB.103.134421} {\bibfield  {journal} {\bibinfo
   {journal} {Phys. Rev. B}\ }\textbf {\bibinfo {volume} {103}},\ \bibinfo
  {pages} {134421} (\bibinfo {year} {2021})}\BibitemShut {NoStop}%
\bibitem [{\citenamefont {Czapkiewicz}\ \emph {et~al.}(2008)\citenamefont
  {Czapkiewicz}, \citenamefont {Stobiecki},\ and\ \citenamefont {van
  Dijken}}]{czapkiewicz2008thermally}%
  \BibitemOpen
  \bibfield  {author} {\bibinfo {author} {\bibfnamefont {M.}~\bibnamefont
  {Czapkiewicz}}, \bibinfo {author} {\bibfnamefont {T.}~\bibnamefont
  {Stobiecki}},\ and\ \bibinfo {author} {\bibfnamefont {S.}~\bibnamefont {van
  Dijken}},\ }\bibfield  {title} {\bibinfo {title} {Thermally activated
  magnetization reversal in exchange-biased [pt/ co] 3/ pt/ ir mn
  multilayers},\ }\href@noop {} {\bibfield  {journal} {\bibinfo  {journal}
  {Physical Review B}\ }\textbf {\bibinfo {volume} {77}},\ \bibinfo {pages}
  {024416} (\bibinfo {year} {2008})}\BibitemShut {NoStop}%
\bibitem [{\citenamefont {Kim}()}]{camley_chapter_2012}%
  \BibitemOpen
  \bibfield  {author} {\bibinfo {author} {\bibfnamefont {J.-V.}\ \bibnamefont
  {Kim}},\ }\bibfield  {title} {\bibinfo {title} {Chapter four - spin-torque
  oscillators}\ }(\bibinfo  {publisher} {Academic Press})\ pp.\ \bibinfo
  {pages} {217--294},\ \bibinfo {note} {{ISSN}: 0081-1947}\BibitemShut
  {NoStop}%
\end{thebibliography}%
\end{document}